\documentclass[iop]{emulateapj}
\usepackage{amsmath}
\begin{document}

\title{{\it WISE} TF: A Mid-infrared, 3.4 $\mu$\lowercase{m} Extension of
  the Tully-Fisher Relation Using \textit{WISE} Photometry}

\author{David J. Lagattuta\altaffilmark{1,2}}
\author{Jeremy R. Mould\altaffilmark{1,2}}
\author{Lister Staveley-Smith\altaffilmark{2,3}}
\author{Tao Hong\altaffilmark{2,3,4}}
\author{Christopher M. Springob\altaffilmark{2,3,5}}
\author{Karen L. Masters\altaffilmark{6,7}}
\author{B\"{a}rbel S. Koribalski\altaffilmark{8}}
\author{D. Heath Jones\altaffilmark{9}}

\shorttitle{The \textit{WISE} Tully Fisher Relation}
\shortauthors{Lagattuta et al.}

\email{dlagattu@astro.swin.edu.au}

\altaffiltext{1}{Centre for Astrophysics \& Supercomputing, Swinburne
  University of Technology, P.O.\ Box 218, Hawthorn, VIC 3122,
  Australia}
\altaffiltext{2}{ARC Centre of Excellence for All-sky Astrophysics
  (CAASTRO)}
\altaffiltext{3}{International Centre for Radio Astronomy Research
  (ICRAR), University of Western Australia, 35 Stirling Highway,
  Crawley, WA 6009, Australia}
\altaffiltext{4}{National Astronomical Observatories, Chinese Academy
  of Sciences, 20A Datun Road, Chaoyang District, Beijing 100012,
  China}
\altaffiltext{5}{Australia Astronomical Observatory, PO Box 915, North
  Ryde, NSW 1670, Australia}
\altaffiltext{6}{Institute for Cosmology and Gravitation, University
  of Portsmouth, Dennis Sciama Building, Burnaby Road, Portsmouth, PO1
  3FX, UK}
\altaffiltext{7}{South East Physics Network, www.sepnet.ac.uk}
\altaffiltext{8}{CSIRO Astronomy and Space Science, Australia
  Telescope National Facility (ATNF) PO Box 76, Epping, NSW 1710,
  Australia}
\altaffiltext{9}{School of Physics, Monash University, Clayton, VIC
  3800, Australia}

\slugcomment{Accepted to The Astrophysical Journal: 9 May 2013}

\begin{abstract}
We present a mid-infrared Tully-Fisher (TF) relation using photometry
from the 3.4 $\mu$m $W1$ band of the \textit{Wide-field Infrared
  Survey Explorer} ({\it WISE}) satellite.  The {\it WISE} TF relation
is formed from 568 galaxies taken from the all-sky 2MASS Tully-Fisher
(2MTF) galaxy catalog, spanning a range of environments including
field, group, and cluster galaxies.  This constitutes the largest
mid-infrared TF relation constructed, to date.  After applying a
number of corrections to galaxy magnitudes and line widths, we measure
a master TF relation given by $M_{\rm corr} = -22.24 -
10.05[\log(W_{\rm corr}) - 2.5]$, with an average dispersion of
$\sigma_{\rm WISE} = 0.686$ magnitudes.  There is some tension between
{\it WISE} TF and a preliminary 3.6 $\mu$m relation, which has a
shallower slope and almost no intrinsic dispersion.  However, our
results agree well with a more recent relation constructed from a
large sample of cluster galaxies.  We additionally compare {\it WISE}
TF to the near-infrared 2MTF template relations, finding a good
agreement between the TF parameters and total dispersions of {\it
  WISE} TF and the 2MTF K-band template.  This fact, coupled with
typical galaxy colors of $(K - W1) \sim 0$, suggests that these two
bands are tracing similar stellar populations, including the older,
centrally-located stars in the galactic bulge which can (for galaxies
with a prominent bulge) dominate the light profile.


\end{abstract}

\section{Introduction}
Scaling relationships play an important role in understanding the
nature and behavior of galaxies.  This is because it is often possible
to infer properties that are difficult to measure (such as stellar
age) from other, more readily available information (such as
metallicity fraction).  While there are a number of scaling relations
currently in use in modern astronomy, such as the Faber-Jackson law
\citep{fj76}, the $D_n$-$\sigma$ relationship \citep{dressler87}, the
fundamental plane \citep{dre87,djo87}, and the bulge black hole mass
relation \citep{fer00}, rotationally dominated (i.e., spiral) galaxies
can be described by the Tully-Fisher (TF) relation.  First measured by
\citet{tf77}, the TF relation measures the correlation between a
galaxy's total luminosity and maximum rotational velocity, finding
that the brighter a galaxy is, the faster it rotates, leading to wider
measured rotation curves.

Originally, the TF relation was measured using B-band luminosities
\citep{tf77}, but a large dispersion around the fitted slope (due
primarily to dust extinction, scattering, and galactic attenuation)
somewhat limited its utility.  Later speculations suggested that
longer-wavelength data, being less susceptible to extinction and
recent star formation, would show a smaller dispersion \citep{aar79},
and this prompted the creation of TF relations in redder bands.  In
particular, there have been several TF relations constructed in both
the optical (e.g., \citealt{bot87,gio97,tul00,ver01,mas06,moc12}) and
the near-infrared (e.g.,
\citealt{aar80,ber94,rot00,mac01,mas08,the07}).  While the specific TF
parameters (slope and intercept of a linear fit) vary from study to
study, the TF scatter does generally decrease as the observed
wavelength increases.

In this work, we continue to extend the redward trend of TF relations,
using luminosity data obtained from the \textit{Wide-field Infrared
  Survey Explorer} ({\it WISE}; \citealt{wri10}) satellite.  Like the
previous 2-Micron All Sky Survey (2MASS; \citealt{skr06}), the {\it
  WISE} mission was designed to image the entire sky in the IR regime.
However, {\it WISE} is equipped with much redder bandpass filters
(3.4, 4.6, 12, and 22 $\mu$m), making it a mid-infrared (mid-IR)
instrument.

At the very longest mid-IR wavelengths, photons from warm dust and gas
dominate galaxy luminosity, rather than direct stellar emission.  Due
to the spatially variable nature of dust in galaxies, TF relations
constructed at these wavelengths may show significantly increased
scatter, as the empirical relationship (originally designed to match
stellar luminosity and rotation) is altered by this extra flux.  In
order to avoid this possibility, we only focus on the shortest {\it
  WISE} bandpass (3.4 $\mu$m), where, after fitting a 300 K blackbody
spectrum to the mid-IR spectral energy distribution of a typical star
forming galaxy (e.g., \citealt{zhe07}), we find that only 1\% to 5\%
of the total emission at 3.4 $\mu$m is due to dust.

By utilizing 3.4 $\mu$m data, we are able to compare our results to
the surprising 3.6 $\mu$m TF relation presented in \citet{fre10}.  In
that work, a mid-IR TF relation was constructed from a sample of eight
galaxies cross-matched between the \textit{Hubble Space Telescope} Key
Project's Cepheid distance sample \citep{sak00} and a catalog of
nearby galaxies \citep{dal07} imaged with the \textit{Spitzer Space
  Telescope}.  (Two of these galaxies, NGC 4725 and NGC 7331, are
included in our final data set.)  Without explicitly correcting for
several systematic effects, the sample showed a remarkably small
dispersion ($\sigma \sim 0.12$ mags), opening up the possibility that
the mid-IR TF relation could be used as a tool for measuring precision
cosmological and distance parameters.  However, as correctly noted in
that paper, a much larger sample of galaxies -- preferably containing
objects with more commonly observed magnitudes -- is needed to verify
(or reject) this result.  Using data acquired from an all-sky
instrument such as {\it WISE}, we are in a prime position to attempt
such a task.

This paper, along with the preliminary work by \citet{fre10} and the
work of \citet{sor12a,sor12b,sor13}, presents the first TF
relationship constructed at mid-infrared wavelengths, and allows us to
probe some empirical properties of galaxies in a regime where the
oldest (assuming inside-out growth of stellar mass; e.g.,
\citealt{per13}), most centrally-located stars make up a slightly more
prominent fraction of the total luminosity \citep{but10}.  In addition
to galaxy properties, however, the TF relation can also be used for
cosmological purposes.  Specifically, the TF relation can be used to
measure peculiar velocities, which can, when taken over a large enough
volume, begin to constrain the dark matter distribution in the Local
Universe.  With a potentially lower TF scatter in the mid-IR, peculiar
velocities can be more robustly determined, ultimately resulting in
better dark matter mapping and an improved understanding of the nature
of the Universe's mass structure.

This paper is organized as follows: In Section 2 we describe the data
products (magnitudes and line widths) used in this work, along with
the initial catalog and specific data cuts used to construct the final
galaxy sample.  In Section 3, we outline our TF analysis, highlighting
several corrections designed to homogenize the galaxy sample.  It is
in this section where we present the final, best-fit {\it WISE} TF relation.
We discuss the results of the TF analysis in Section 4.  Specifically,
we compare the parameters of {\it WISE} TF to other near- and mid-IR TF
relations and identify potential sources of intrinsic TF scatter at
mid-IR wavelengths.  We briefly conclude in Section 5.  Additionally,
technical details testing the accuracy of the peculiar velocity
corrections, along with measurements showing the effects of the
corrections on the final TF relation, are presented in an Appendix.

Throughout this work, we assume a Hubble parameter $H_0 = 100 ~h$ km
s$^{-1}$ Mpc $^{-1}$.  All magnitudes are calculated in the Vega
system.

\section{Data}
\label{data}
We select the initial TF candidate galaxies by cross-matching
magnitudes from the {\it WISE} $W1$ (3.4 micron) all-sky data release
catalog with rotational velocity (\ion{H}{1} line width) measurements
from the 2MASS Tully Fisher (2MTF; \citealt{mas08}) all-sky galaxy
catalog \citep{mas08,hon13}.  While the complete 2MTF catalog contains
line width measurements acquired from archival sources -- the Cornell
\ion{H}{1} digital archive \citep{spr05} and the HYPERLEDA database
\citep{the98,the05,pat03} -- and new observational data from the Green
Bank Telescope and the Parks Radio Telescope \citep{hon13}, we
construct the {\it WISE} TF relation using only the publicly available
line width data (i.e., the Cornell and HYPERLEDA samples).  Although
there are selection effects embedded in the overall archival catalogs
(due to the heterogeneous nature of how they were compiled), the
subset of the data comprising the 2MTF sample is specifically designed
to be complete down to magnitude $K = 11.5$, minimizing any potential
selection bias.  We explicitly exclude galaxies in a related catalog,
the 2MTF ``galaxy template'' sample \citep{mas08} from our final data
set, as this sample will be analyzed separately in a companion paper.
Because the remaining ``non-template'' 2MTF galaxies cover a larger
area of the sky, include a wider range of environments (from clusters,
to groups, to field galaxies), and span a similar redshift range ($0 <
z < 0.05$), removing the galaxy template sample from our analysis will
not bias the final results.

In the {\it WISE} catalog, magnitudes of resolved 2MASS galaxies are
measured using an elliptical aperture designed to mimic the size and
shape of a given galaxy.  This is achieved by convolving the galaxy's
shape, as measured in the 2MASS Extended Source catalog \citep{skr06},
with the {\it WISE} $W1$ point spread function (PSF).  This particular
magnitude, called \textit{w1gmag}, captures more light than the
standard, profile-fit magnitude \textit{w1mpro} (optimized for
measuring point-sources) and is less sensitive to systematics from
inclination effects.  While a \textit{w1gmag} magnitude can
underestimate the true magnitude by as much as
20\%\footnote{http://wise2.ipac.caltech.edu/docs/release/allsky/expsup/sec6\_3e.html},
this is a significant improvement over \textit{w1mpro}, which
typically underestimates extended source flux by 60-80\%.  This
underestimation is encoded in the final magnitude uncertainty.

We note that alternatives to {\it w1gmag} are also in development (see
e.g, \citealt{jar12,jar13}) which may provide improved magnitude
estimates.  In particular, an isophotal curve-of-growth analysis
(combined with the careful removal of foreground stellar light) yields
$W1$ galaxy magnitudes that are on average $0.32 \pm 0.14$ mags
brighter than their {\it w1gmag} counterparts (T. Jarrett and
M. Bilicki, private communication).  We can therefore revisit the
issue of galaxy photometry when these alternative catalogs are
released publicly.  However, with a potential scatter of only $\sim
0.14$ mags relative to the \textit{w1gmag} values, these updated
measurements will not significantly affect the TF slope or dispersion
parameters in our analysis (Section \ref{TF_construction}).

After the initial cross-matching, We begin with a set of 1929
galaxies.  However, in order to improve the quality of the sample we
apply a number of data cuts.  Only galaxies with the highest line
width quality flags (``G'' from the \ion{H}{1} archive and ``A'' from
the HYPERLEDA data; see \citealt{spr05} and \citealt{the05},
respectively for flag descriptions) are kept, as these are most
accurate measurements for TF purposes.  To reduce the intrinsic
scatter of the line width measurement, we apply a signal-to-noise
cutoff of $\sigma_{\rm width}$/width $< 0.1$, and also reject galaxies
with an inclination angle less than $45^{\degr}$.  Inclination angles
are derived from the axis ratio of each galaxy, but rather than
measuring these quantities directly from the {\it WISE} catalog, we instead
use the optical (V-band) axis ratios presented in the NASA/IPAC
Extragalactic Database (NED)\footnote{http://ned.ipac.caltech.edu/}.
This is done so that the axis ratio is optimized to measure the bluer
disk light rather than the redder bulge light, which often
underestimates ellipticity.

A morphology cut is also applied to the data, to ensure that we only
select rotationally-supported galaxies in the final
sample. Specifically, we select galaxies classified between type S0 (T $=0$)
and Sd (T $=9$), using the de Vaucouleurs T-type classification scheme.
Morphology information is taken from the CfA Redshift Catalog (ZCAT;
\citealt{huc12}).  

After applying all cuts, we are left with 568 galaxies in the final
sample.  While the overall number has been dramatically reduced, the
physical characteristics of the remaining galaxies are not
significantly different from the original population.  In particular,
shapes of galaxy magnitude, line width, and recessional velocity
distributions remain the same before and after the cuts are applied,
and the final sample still covers a large area across the sky.
Because of this, we are confident that the reduced data set is not
fundamentally different from the full set, and analysis of the final
galaxy sample will yield robust results.

\begin{figure}
\begin{center}
\centerline{
\includegraphics[width=7.8cm]{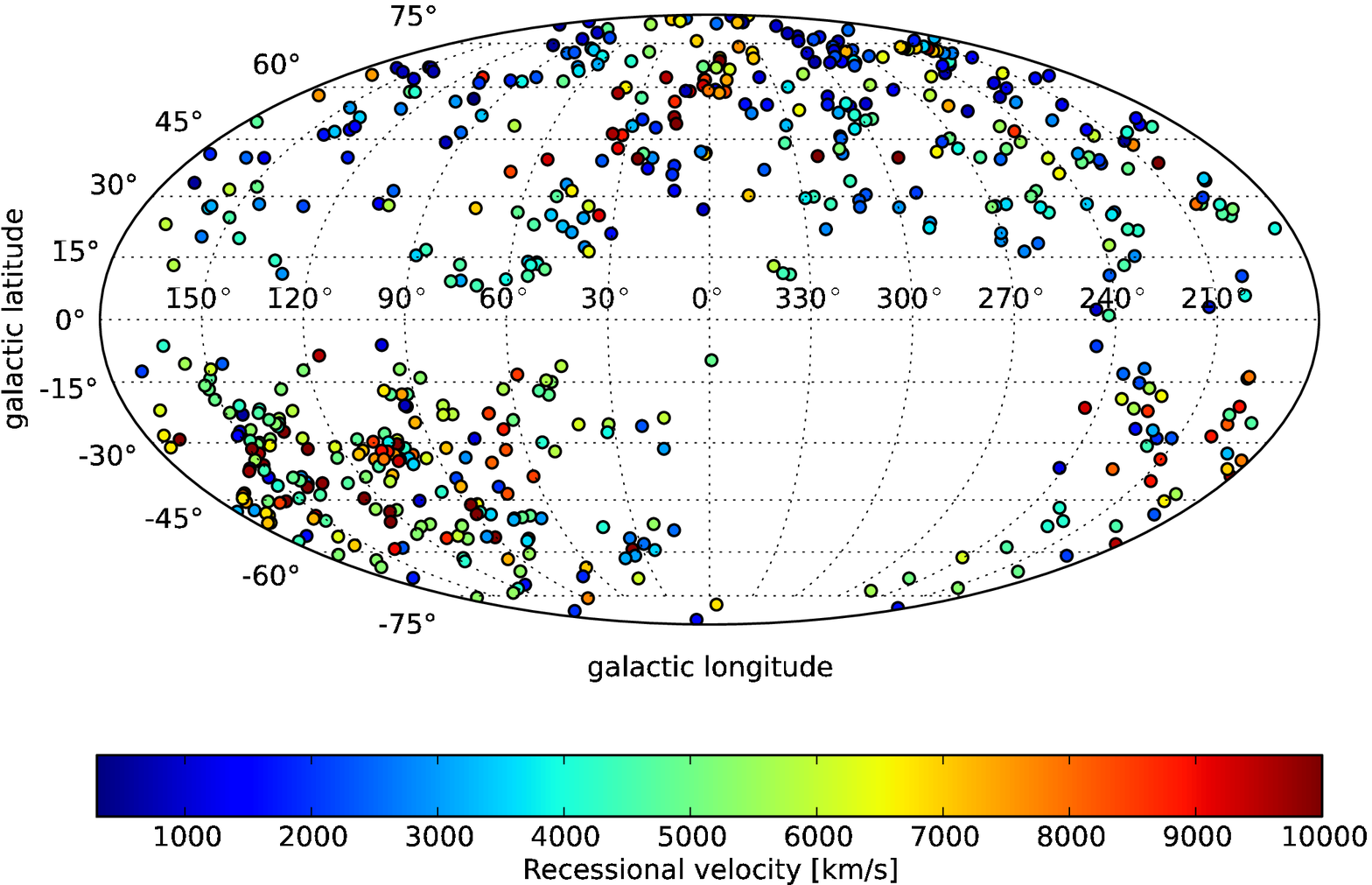}}\hfill~
\centerline{
\includegraphics[width=7.8cm]{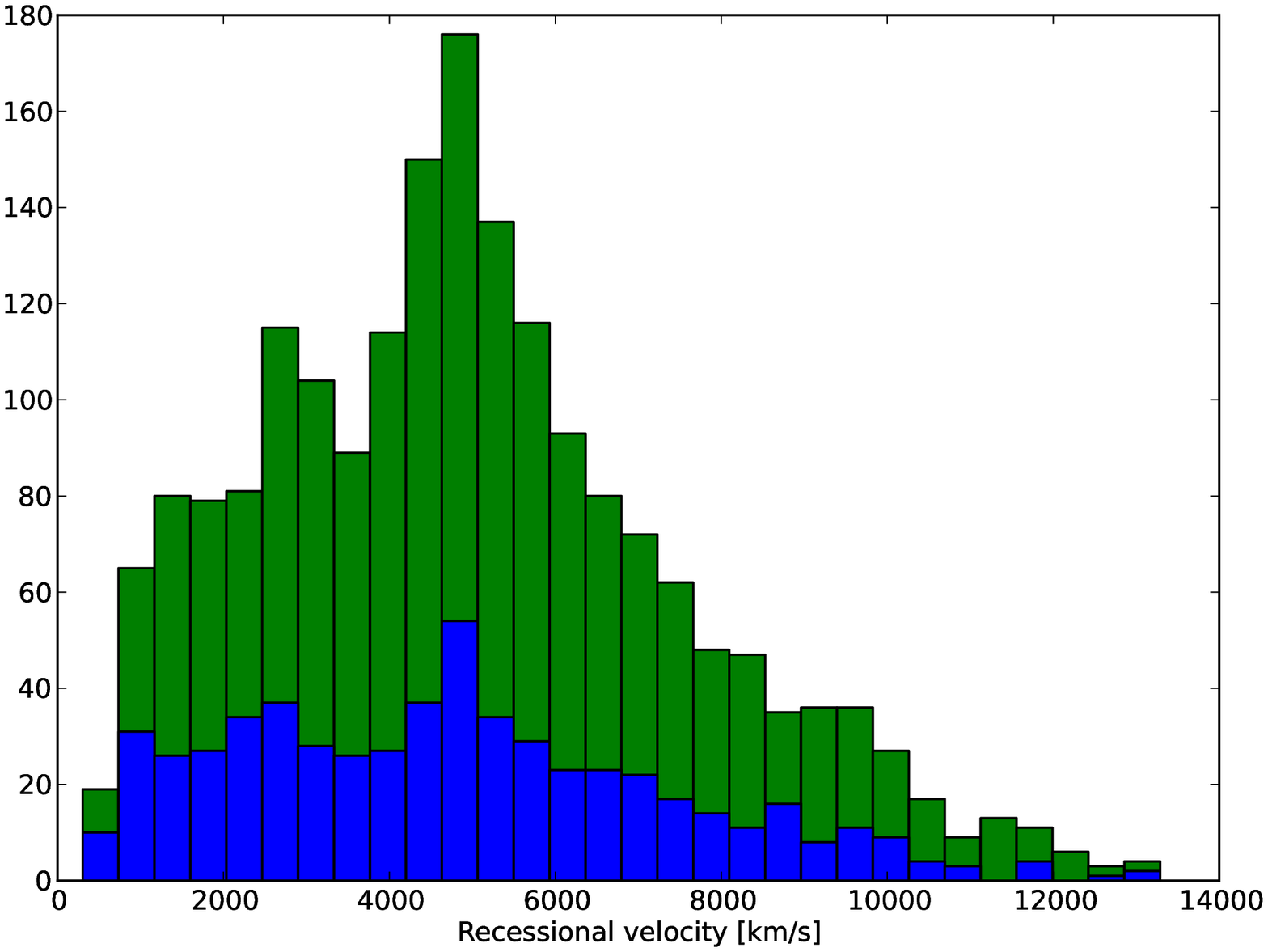}}
\end{center}
\caption{Coordinates and recessional velocities of the final {\it
    WISE} TF galaxy sample.  {\bf Top:} A scatterplot of galaxy
  coordinates, using an equal-area Mollweide projection.  The data gap
  seen around the Equatorial South Pole is due to a lack of publicly
  available, calibrated 2MTF line widths.  This region will be filled
  in as the remaining portions of the 2MTF catalog are published.  The
  recessional velocity (redshift) of a given galaxy is indicated by
  color, though we apply an artificial color cutoff at v = 10000 km
  s$^{-1}$ to improve the contrast of the distribution.  {\bf Bottom:}
  A velocity histogram showing the redshift distribution of our galaxy
  sample both before applying the specific cuts in Section \ref{data}
  (green) and after(blue).  The distribution peaks at v = 5000 km
  s$^{-1}$ ($z \sim 0.015$), but galaxies continue out to v = 14000 km
  s$^{-1}$ ($z \sim 0.045$).\label{fig:galaxy_distrib}}
\end{figure}

A map of the positions of sample galaxies (in Galactic coordinates),
along with a redshift distribution histogram can be seen in Figure
\ref{fig:galaxy_distrib}.  Since, as mentioned above, we do not use
the 2MTF galaxy sample observed with the Parkes telescope
\citep{hon13}, our sample is artificially cut off at $-45^\circ$
declination.  In the lower panel, both the pre-cut (green) and
post-cut (blue) redshift distributions are shown, highlighting the
fact that our cuts do not significantly bias the physical properties
of the galaxy sample.  The observed redshift of the sample extends to
14000 km s$^{-1}$ ($z \sim 0.045$).

%--------------------------------------------------------------------
%--------------------------------------------------------------------

\section{Template Construction and Analysis}

Before attempting to fit a TF relation to our galaxy sample, we first
apply a number of corrections to ensure homogeneity.  We now briefly
describe each of these corrections, and our specific methods for
implementing them.

\subsection{Luminosity Corrections}

First, we apply several photometric corrections to the observed
apparent magnitudes of the sample.  In all cases, the distance modulus
required for converting a galaxy's apparent magnitude into the
absolute frame is the largest correction, but there are a number of
smaller corrections necessary as well.  These include adjustments for
peculiar velocity ($\mu_{\rm H}$), Galactic extinction ($A_{3.6}$), and
internal absorption ($\Delta m_{\rm Int}$).  We now describe each of
these processes.

\subsubsection{Peculiar Velocity}
\label{vpec}
In the nearby Universe ($z < 0.1$), the peculiar velocity of a galaxy
(i.e. additional speed and direction relative to the Hubble flow), is
typically on the order of 200 km s$^{-1}$ \citep{pee76}.  Over the
distance range of the {\it WISE} TF galaxy sample, this translates to
a (cosmological) redshift uncertainty between 1.4 and 100\%, giving
rise to an additional 0.03 to 1.5 magnitudes of scatter in the TF
relation.  Removing the effects of this peculiar motion should,
therefore, reduce the observed scatter.  In this work, we achieve this
by applying the peculiar velocity model of \citet{erd06} to each
galaxy.

Since the \citet{erd06} model is coarsely sampled, with peculiar
velocity values only reported at regular (5 Mpc x 5 Mpc x 5 Mpc)
intervals in Cartesian Supergalactic space, we rely on interpolation
to improve the model's utility.  In particular we apply a 3rd-order,
bicubic spline interpolation scheme to the model grid so that we are
able to apply a unique peculiar velocity value at the coordinates of
each sample galaxy.  Systematic tests assessing the accuracy of our
spline interpolation scheme can be seen in the Appendix.

The model provides an estimate of total velocity ($v_{\rm T}$) as a
function of Hubble flow velocity ($v_{\rm H}$) for any supergalactic
(sgl, sgb) position.  We therefore interpolate over the inverse
function, $v_{\rm H}(v_{\rm T})$, (assuming $v_{\rm T} = v_{\rm obs}
\equiv c z$, the galaxy's observed redshift velocity) to remove the
effects of peculiar velocity.  The CMB reference frame is used
throughout this process.

While, in general, it is not always possible to uniquely invert the
total velocity function due to complications from the ``triple-value
problem'' (see e.g., \citealt{dav83,mar98}), we find $v_{\rm T}$ to be
monotonic with redshift for all galaxies in our sample, resulting in
an unambiguous Hubble flow velocity measurement every time.  The
$v_{\rm H}$ values can then be used to derive a corrected distance
modulus, according to the equation:

\begin{equation}
\label{eqn:vpec}
\mu_{\rm H} = -5 \log \left[ \frac{v_{\rm H}}{100h} \right ] - 25
\end{equation}

\subsubsection{Extinction}
In addition to peculiar velocity modifications, we also correct the
{\it WISE} photometry for the effects of dust extinction, due to both
Milky Way reddening and internal absorption from the galaxies
themselves.

To correct for Galactic extinction, we apply the dust models of
\citet{sch98}, obtaining reddening values for each galaxy from the
NASA/IRSA dust web
service\footnote{http://irsa.ipac.caltech.edu/applications/DUST/}.
Since the web service only offers corrections for a limited number of
wavebands, we adopt the reddening correction associated with the
3.6-micron IRAC-1 channel ($A_{3.6} = 0.179\cdot E(B-V)$), as this
wavelength is the closest to that of the 3.4 micron {\it WISE} W1 data
we use.  In general, extinction due to Galactic reddening is small
($\Delta~m_{\rm Gal} < 0.05$ mag), however, the correction is still
included for completeness sake.

We must also account for internal extinction by a galaxy's
interstellar medium, an effect which increases with inclination angle.
For this correction, we use the same technique described in
\citet{mas03}: the modification takes the form of a broken power-law
model with a shallow slope for low-to-moderate-inclination galaxies,
and a steeper slope for high-inclination (nearly edge-on) galaxies.
We specifically choose the broken power-law correction rather than a
purely linear one \citep[e.g.,][]{gio94,tul98} to better match the
redder near-IR colors of high-inclination ($\log(a/b) > 0.5)$ galaxies
\citep{mas03}.  To correct the {\it WISE} galaxy sample, we start with
the 2MTF K-band model from \citet{mas08}, but we scale both slopes to
the W1 band by a 1/$\lambda$ scaling law.  This serves to flatten both
the low-end and high-end slopes by nearly a factor of two
($\lambda_{\rm scaling} = 2.2{\mu}$m/$3.4{\mu}$m), but this is
expected, as dust absorption (both internal and external) is much less
prevalent as wavelength increases.  Like the Galactic extinction
mentioned above, internal extinction corrections are also usually
small ($\Delta~m_{\rm Int} < 0.05$ mag), though some of the steeply
inclined galaxies on our sample (inclination $> 75^{\degr}$) can show
as much as a 0.2 mag difference.

\subsubsection{The full correction}
After calculating all of the luminosity corrections, we apply them to
the observed magnitudes, according to the equation:

\begin{equation}
\label{eqn:tf_mag}
M_{\rm corr} = m_{\rm obs} + \mu_{\rm H} - \Delta{m_{\rm Int}} - A_{3.6}
\end{equation}

These corrected magnitudes are then used in the TF relation.

\subsection{Line Width Corrections}

Like the luminosity, rotation widths must also be corrected before a
TF relation can be fit.  In particular, we must account for systematic
effects between different telescopes, line width broadening due to
cosmology, turbulent motion in the disks of galaxies, and inclination
of the galaxies themselves.  We now briefly describe these
corrections.

\subsubsection{Instrumental and Cosmological Effects}
Since our line width data is taken from publicly available archives,
many galaxies (especially those found in the Cornell archive) have
already been corrected for both instrumental and cosmological effects
(see \citet{spr05}).  If these corrections are not already available, we
calculate them ourselves using other methods, according to the equation

\begin{equation}
\label{eqn:cosmo_inst_corr}
W_{\rm IC} = \frac{W_{\rm obs} - \Delta_s}{1 + z}
\end{equation}
where $\Delta_s$ represents the systematic instrumental correction,
and the $1+z$ term corrects for cosmological broadening.

\subsubsection{Turbulent motion}
Based on the turbulent motion simulations presented in \citet{spr05},
we apply a constant correction of $\Delta_t = 6.5$ km s$^{-1}$ to
account for potential turbulent motion broadening.  This shifts the
entire TF relation to slightly lower line widths, decreasing the
intercept of the fit by a small amount.

\subsubsection{Galaxy Inclination}
In order to further normalize our line width data, we also apply a
correction for galaxy inclination.  This is done using the standard
$1/\sin(\theta_i)$ broadening correction, where the inclination angle
$\theta_i$ is again calculated from the galaxy's optical axis ratio.

\subsubsection{The full correction}

Just as before, we apply all of the above corrections to the measured
line widths for the TF analysis, according to the following equation:

\begin{equation}
\label{eqn:tf_widths}
W_{\rm corr} = \frac{W_{\rm IC} - \Delta_t}{\sin(i)}
\end{equation}

\subsection{Constructing the Tully Fisher Relation}
\label{TF_construction}
\subsubsection{Initial Construction}
\label{initial_TF}
\begin{figure}
\plotone{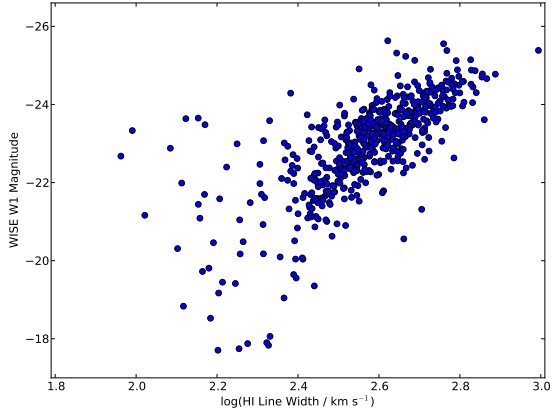}
\caption{First look at the {\it WISE} 3.4 micron Tully Fisher
  relation, plotting total $W1$ magnitude against the logarithm of
  \ion{H}{1} line width.  We see a positive correlation between
  magnitude and line width, and that the dispersion on the TF relation
  decreases towards the bright end.  \label{fig:TF_initial}}
\end{figure}

After applying the corrections to galaxy magnitudes and line widths,
we are able to create a TF relation with the data.  A first look at
this relation is shown in Figure \ref{fig:TF_initial}.
%--------------------------------------------------------------------
%--------------------------------------------------------------------
Like TF relations in other bands, the {\it WISE} TF relation shows a
positive correlation between galaxy magnitude and line width, with an
average dispersion that increases towards the faint end.  To better
quantify this behavior, we fit a line to the data, using a
Levenberg-Marquardt least squares optimization routine to determine
the best-fit slope and intercept parameters. The minimization itself
is applied to the bivariate $\chi^2$ function given by:

\begin{equation}
\label{eqn:chisqr_fit}
\chi^2 = \sum_{i}^{N} \frac{\left(a\cdot W_{{\rm corr}, i} + b - M_{{\rm corr}, i} \right)^2}{(a \cdot
  \sigma_{W, i})^2 + (\sigma_{M, i})^2}
\end{equation}
where $a$ and $b$ are the model slope and offset, respectively,
$W_{\rm corr}$ and $M_{\rm corr}$ are the line widths and magnitudes of
the sample galaxies, and $\sigma_{W}$ and $\sigma_{M}$ are the
uncertainties on the measured data.  These uncertainties are
calculated by adding the intrinsic measurement uncertainties of the
data in quadrature with the errors propagated from the corrections
applied in the previous section.

\begin{figure*}
\plottwo{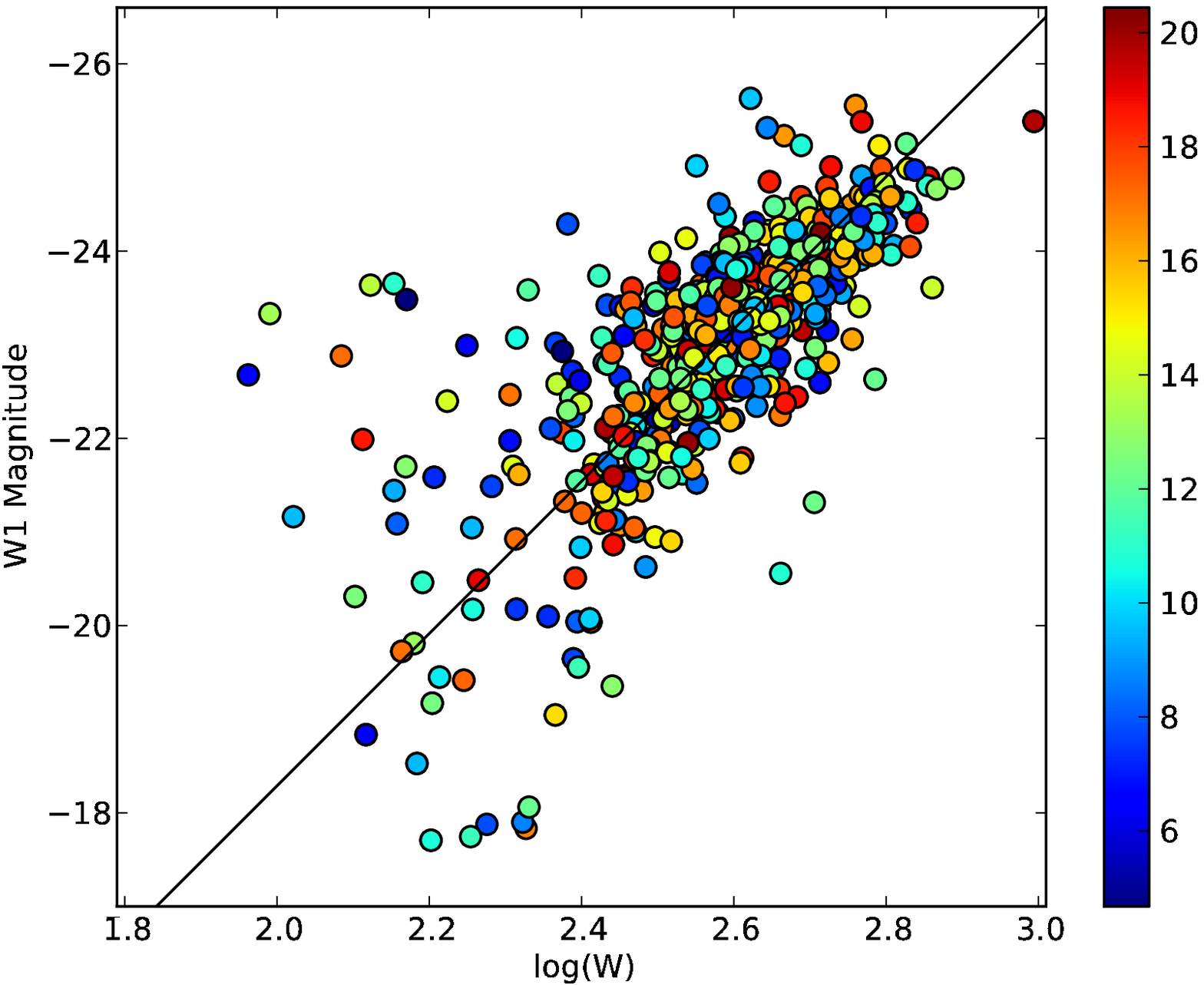}{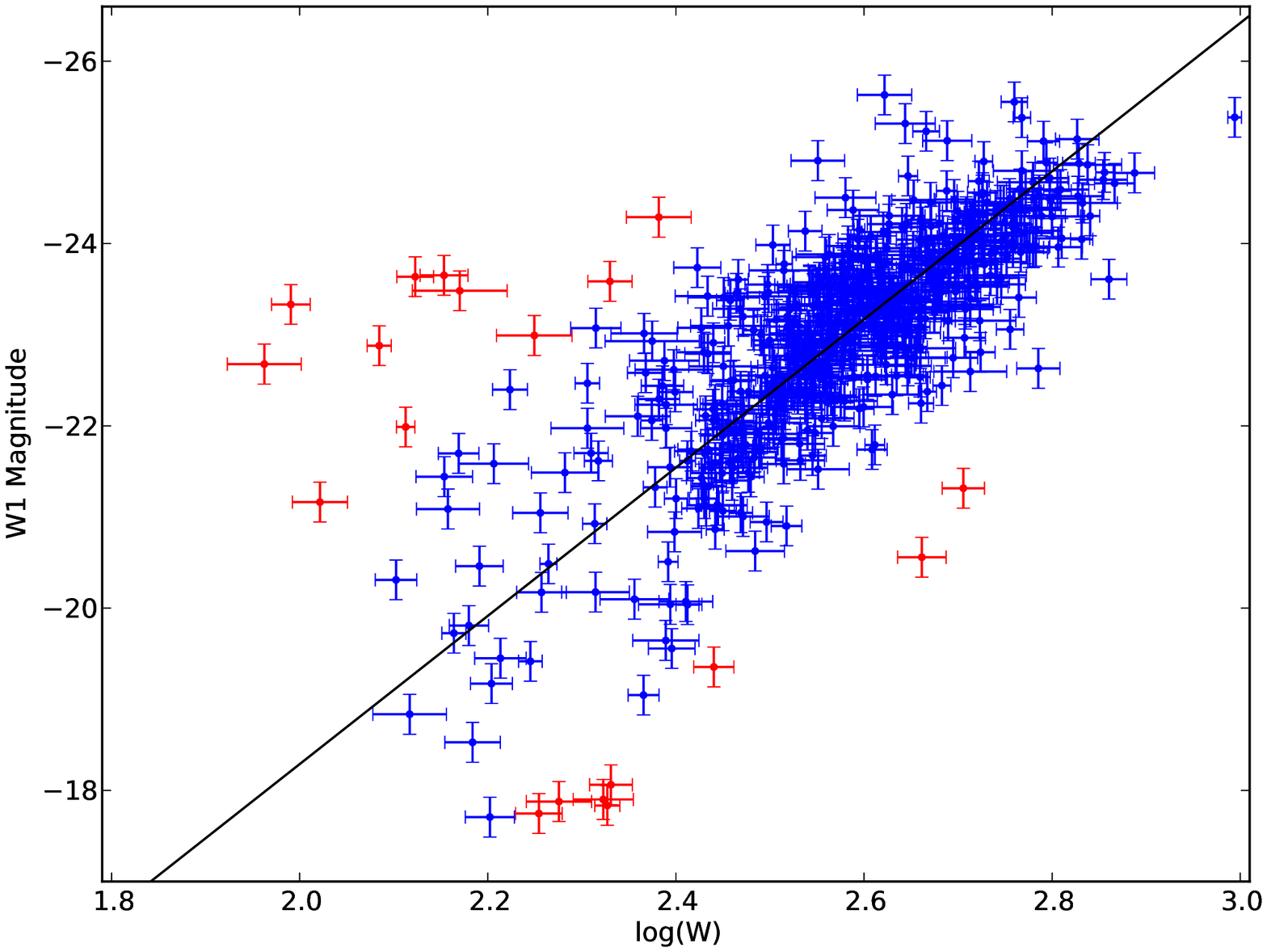}
\caption{Best-fit TF relation (solid black line) of the {\it WISE}
  galaxy sample.  Left: Best-fit line overlaid on the galaxy data.
  The color of a data point represents that galaxy's specific weight
  in the least-squares fit.  Right: Same data/TF relation overlay, but
  with error bars representing the relative contribution of line width
  and magnitude uncertainties to the error budget.  In this panel,
  galaxies removed from the final sample by sigma clipping are
  indicated in red.
  \label{fig:TF_fits}}
\end{figure*}

To prevent outliers from biasing the results, we repeat the fitting
process several times, applying a 3.5$\sigma$ sigma clip to the data
after each iteration.  If after a particular iteration no galaxies are
clipped from the sample (typically after the 3rd or 4th fit), we
consider the fit to have converged and treat the final parameters as
our best-fit TF slope and intercept.  Examples of the fit can be seen
in Figure \ref{fig:TF_fits}.

%--------------------------------------------------------------------
%--------------------------------------------------------------------

For the initial {\it WISE} TF relation, we find a best-fit slope of $a
= -8.13$ and a best-fit intercept of $b = 2.02$. To characterize
uncertainties on the parameters, we analyze 1000 bootstrap resamplings
of the data, measuring an average variation of $\sigma_a = 0.6$ and
$\sigma_b = 1.8$ for the slope and intercept, respectively.

As a check on these parameters, we also fit the data with an inverse
TF relation (assuming $W_{\rm corr}$ as the dependent variable), given
by
\begin{equation}
W_{\rm corr} = M_{\rm corr}/a_{\rm inv} + b_{\rm inv}
\end{equation}
Like before, $a_{\rm inv}$ and $b_{\rm inv}$ are the best fit slope
and intercept, respectively.  To obtain the best fit, we use a
modified $\chi^2$ function:
\begin{equation}
\label{eqn:chisqr_fit}
\chi^2_{\rm inv} = \sum_{i}^{N} \frac{\left(M_{{\rm corr}, i}/a_{\rm
    inv} + b_{inv} - W_{{\rm corr}, i} \right)^2}{(\sigma_{W, i})^2 +
  (\sigma_{M, i}/a_{\rm inv})^2}
\end{equation}
but maintain the 3.5$\sigma$ clip.  The resulting best-fit parameters
remain unchanged with this optimization scheme (i.e., $a_{\rm inv} =
-8.13$, $b_{\rm inv} = 2.02$), and thus we focus only on the bivariate
relation in the remaining analysis.

\subsubsection{Correcting for galaxy morphology}
\label{morphology}

\begin{figure}
\plotone{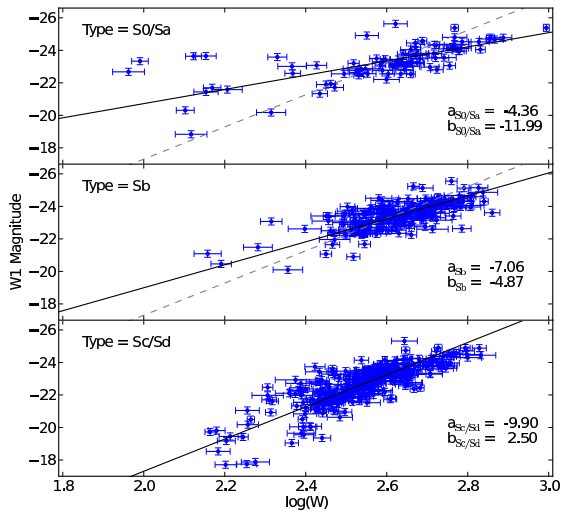}
\caption{TF relations fit to three different morphological types:
  S0/Sa, Sb, and Sc/Sd.  The earlier-type spiral groups clearly have a
  shallower TF slope.  In each panel, the best-fit slope is
  represented by a solid black line.  In the S0/Sa and Sb panels, the
  best fit line from the Sc/Sd morphological group is represented by a
  gray dashed line.\label{fig:morph_subgroup}}
\end{figure}

After initially constructing {\it WISE} TF (Figure
\ref{fig:TF_initial}) we notice a broken power-law feature in the
data, occurring at $\log (W_{\rm corr}) = 2.6$.  \citet{gio97} and
\citet{mas06} both found slight morphological dependencies on TF
parameters, with \citet{gio97} showing that earlier type spirals (S0
through Sb) have a slightly brighter TF intercept, and \citet{mas06}
finding that these spirals have both brighter intercepts and
intrinsically shallower slopes.  It is possible, therefore, that this
morphological dependence may (at least partially) explain the break
that we see.  To investigate this, we separate our galaxy sample
(including those galaxies rejected by the sigma-clip) into three
distinct morphological categories: an S0/Sa group, an Sb group, and an
Sc/Sd group, and fit a separate TF relation to each set.  These fits
can be seen in Figure \ref{fig:morph_subgroup}, and as expected, the
S0/Sa and Sb galaxy samples do indeed have a shallower slope than the
Sc/Sd subsample.  In particular, we find slopes of $a_{S0/Sa} =
-4.36$, $a_{Sb} = -7.06$, and $a_{Sc/Sd} = -9.90$.  These slope
discrepancies are larger than those measured by \citet{gio97} in the
optical, but are comparable to the near-IR slope discrepancies seen in
the 2MTF templates \citep{mas08}.  Furthermore, since each group
occupies a distinct region of TF phase space, with the S0/Sa and Sb
groups laying mainly above $\log (W_{\rm corr}) = 2.6$ and the Sc/Sd
group positioned mainly below $\log (W_{\rm corr}) = 2.6$, the
morphology dependence does appear to explain why the broken power-law
feature is so distinct in Figure \ref{fig:TF_initial}.

%--------------------------------------------------------------------
%--------------------------------------------------------------------

Following \citet{mas08}, we apply an additional magnitude correction
to S0/Sa and Sb galaxies, in order to normalize the data into a single
Sc/Sd-equivalent TF relation.  By doing this, we further reduce the
dispersion on the {\it WISE} TF relation, and can more easily compare
it to the 2MTF templates, which are normalized to Sc.  We first fit
new TF relations to the S0/Sa and Sb galaxy samples, but force the
slopes to match the Sc/Sd sample slope.  Differences between the Sc/Sd
TF intercept and the new, fixed TF relation intercepts ($\Delta
b_{S0/Sa} = 0.21$; $\Delta b_{Sb} = 0.10$) are then added to the
magnitudes of a given morphological type.  To correct the ``tilt'' of
the earlier-type spirals, we also include a line-width-dependent
magnitude correction, given generally by $\Delta M = \Delta a \times
[\log(W) - 2.5]$, where $\Delta a$ is the difference between the S0/Sa
or Sb slopes and the Sc/Sd slope presented above.

\begin{figure*}
\plotone{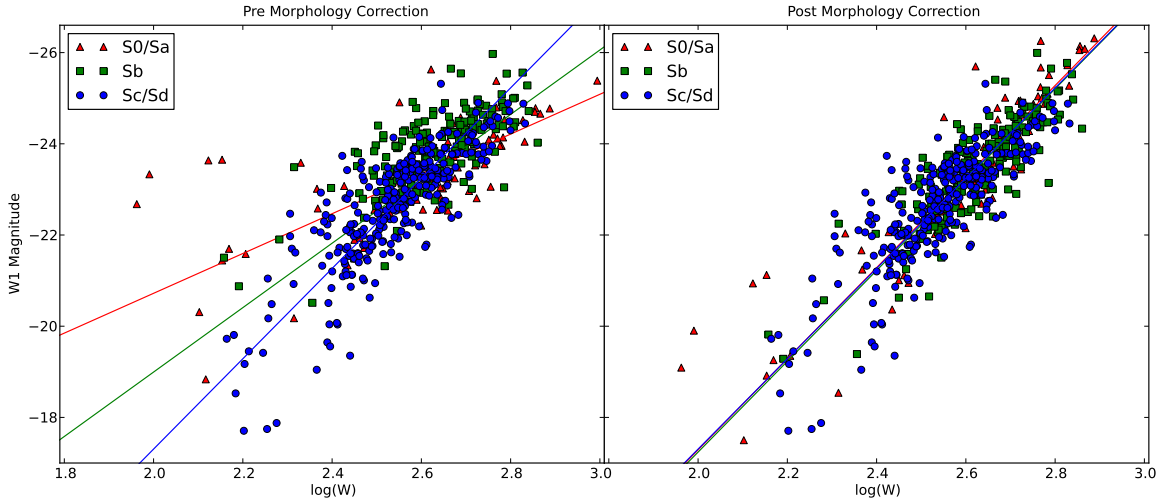}
\caption{Effects of morphology correction on the {\it WISE} TF
  relation.  {\bf Left:} The three distinct morphological categories,
  S0/Sa (red triangles), Sb (green squares), and Sc/Sd (blue circles)
  plotted together before magnitude corrections are applied to the
  earlier-type groups.  We note that the three groups coincide at
  $\log(\rm W) = 2.6$, the location of the sharp power-law break.
  {\bf Right:} The same three morphological types after applying the
  magnitude corrections.  All galaxies are now normalized to the Sc/Sd
  TF relation, and can be fit together to create a ``master'' TF
  relation.  \label{fig:morph_correct}}
\end{figure*}

After applying both the constant and line-width-dependent offsets to
the S0/Sa and Sb galaxy samples, we find a secondary offset between
the different morphological groups ($\Delta b_{S0/Sa, \rm new} = 0.4$;
$\Delta b_{Sb, \rm new} = 0.20$).  We therefore manually correct the
galaxy magnitudes to account for these offsets, without re-fitting a
TF relation.  The final magnitude corrections for
morphology are given by
\begin{align}
\Delta M_{S0/Sa} &= -0.61 - 5.54 \left[\log(\rm W_{corr}) - 2.5 \right]\\
\Delta M_{Sb}~~~~ &= -0.30 - 2.84 \left[\log(\rm W_{corr}) - 2.5 \right]
\end{align}
and the effects of these corrections can be seen in Figure
\ref{fig:morph_correct}.

%--------------------------------------------------------------------
%--------------------------------------------------------------------

\begin{figure}
\plotone{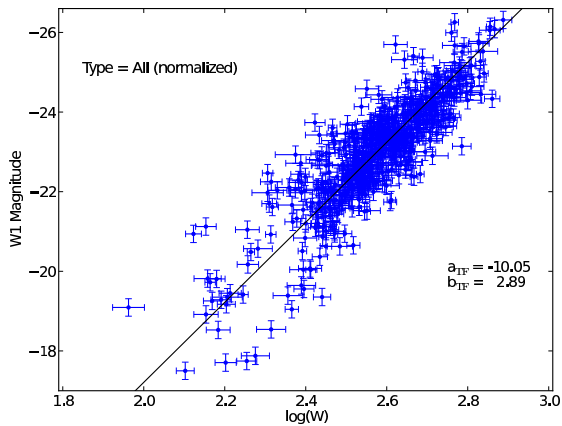}
\caption{The {\it WISE} TF relation after all galaxies are normalized
  to an Sc/Sd morphology.  While the slope of the normalized relation
  increases and the intercept gets dimmer, the overall dispersion of
  the relationship does not significantly
  change. \label{fig:TF_normalized}}
\end{figure}

With all galaxies normalized to an Sc template, we fit a new TF
relation to the entire {\it WISE} TF galaxy sample (Figure
\ref{fig:TF_normalized}), using the same bivariate fitting and
sigma-clipping procedure outlined previously.  The normalized relation
shows a steeper overall slope ($a_{\rm TF} = -10.05$) and fainter
intercept ($b_{\rm TF} = 2.89$) when compared to the unnormalized
galaxy sample.  We take these values to be our final, best-fit
parameters, and use them to create a master {\it WISE} TF relation:

\begin{equation}
M_{\rm corr} = -22.24 - 10.05[\log(W_{\rm corr}) - 2.5]
\end{equation}
The data used to construct this final relation, including the observed
galaxy magnitudes and line widths, the corrections applied to the
observed quantities, and final data sigma-clipping can be seen in
Table \ref{tbl:TF_data}.

To measure the total dispersion for the master {\it WISE} TF, we
calculate the weighted standard deviation around its fitted slope.  We
employ an inverse variance weighting scheme, taking the total variance
for a given galaxy to be the quadrature sum of the uncertainties in
line width and magnitude.  For our data, we find a total dispersion of
$\sigma_{\rm WISE} = 0.686$ magnitudes.

%--------------------------------------------------------------------
%--------------------------------------------------------------------

\section{Discussion}

\subsection{Comparing {\it WISE} TF to other TF relations}

\begin{figure}
\plotone{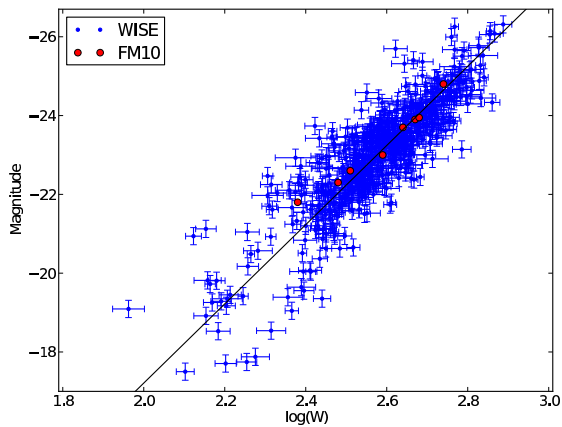}
\caption{The \citet{fre10} TF 3.6 $\mu$m galaxy sample over-plotted on
  the {\it WISE} TF relation (after adjusting the magnitude
  zeropoints).  The sample is fully consistent with our own set of
  galaxies, so it is likely a case of small-number statistics and
  chance alignment that gives rise to the anomalously low TF
  scatter. \label{fig:WTF_vs_FM}}
\end{figure}

From Figure \ref{fig:TF_normalized}, it is clear that the mid-IR TF
relation does not maintain the near-zero scatter presented in Figure
7d of \citet{fre10}, though this is not entirely surprising.  Since
the sample used in \citet{fre10} is quite small (consisting of only 8
galaxies), it is possible that small-number statistics and chance
alignment conspire to create the unusual result -- a fact that
Freedman \& Madore themselves point out.  By including a much larger
sample of galaxies, as we have, one can gain a more complete picture
of mid-IR TF magnitudes, leading to the larger dispersion estimates we
see in our sample.  In fact, over-plotting the \citet{fre10} sample on
to the {\it WISE} TF relation (after adjusting the sample's
zeropoint), we find that they are perfectly consistent with many of
our own galaxies (Figure \ref{fig:WTF_vs_FM}).

%--------------------------------------------------------------------
%--------------------------------------------------------------------

The \citet{sor13} mid-IR TF relation provides a better point of
comparison for {\it WISE} TF.  Also using \textit{Spitzer} 3.6 $\mu$m
photometry, \citet{sor13} construct a TF relation using $\sim$200
galaxies found in large clusters.  While the photometric corrections
they apply to their galaxies are slightly different from our own (they
include small corrections for Doppler shifting and PSF diffusion, and
do not include corrections for peculiar velocity or morphology), and
the analysis technique is not the same (in particular they only
consider errors on line widths, rather than the bivariate fitting
scheme we employ), we find that the final TF parameters agree well
with {\it WISE} TF: $M_{3.6} = -22.84 -9.47[\log(W_{3.6}) - 2.5]$,
where the 3.6 $\mu$m zeropoint has been corrected from AB to Vega
magnitudes.  The total dispersion of this TF relation is $\sigma_{3.6}
= 0.49$ mags, slightly lower that the value we find for {\it WISE} TF,
but still much closer than that presented in \citet{fre10}.

However, in addition to this ``raw'' TF relation, \citet{sor13} apply
an extra (i - [3.6]) color correction to their galaxies in order to
reduce the total scatter, showing that redder galaxies tend to lie
above their mean TF relation and bluer galaxies lie below (see Figure
7 in \citealt{sor13}).  This does reduce the scatter ($\sigma_{\rm
  3.6,C} = 0.44$ mags) but also results in a shallower slope ($a_{\rm
  3.6,C} = -9.13$).  We do not apply this correction to our sample,
but doing so may reduce the overall dispersion by up to 10\%, based on
the reduction to the 3.6 $\mu$m relation.

\begin{figure}
\plotone{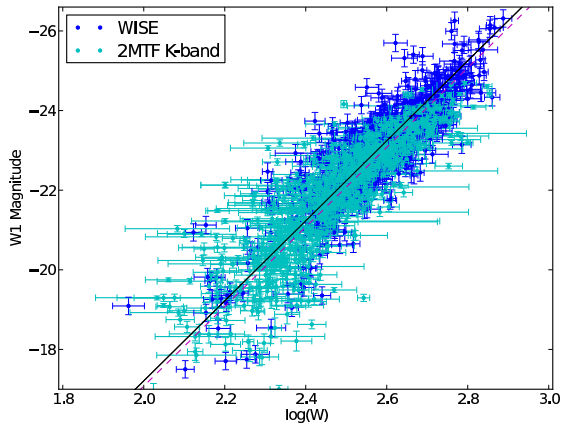}
\caption{A comparison of the {\it WISE} TF relation (blue points) and
  the K-band 2MTF relation (cyan points) of \citet{mas08}.  The solid
  black line again shows the best-fit TF parameters for {\it WISE} TF,
  while the dashed magenta line shows the same for the $K$-band.  The
  two relations are described by remarkably similar parameters,
  suggesting that both $K$- and $W1$-band observations are probing
  very similar stellar populations. \label{fig:WISE_vs_K}}
\end{figure}

Since our galaxy data is drawn from the 2MTF catalog, it is also
interesting to see how the {\it WISE} TF relation compares to the 2MTF
templates.  A comparison between the normalized {\it WISE} TF relation
and the K-band 2MTF relation from \citet{mas08} is shown in Figure
\ref{fig:WISE_vs_K}.  For all three important TF parameters, the two
relations are remarkably similar:

\begin{enumerate}
\item Slope: ($a_{\rm WISE} = -10.05$, $a_{\rm K} = -10.01$)

\item Zeropoint: ($b_{\rm WISE} = -22.24$, $b_{\rm K} = -22.03$)

\item Total dispersion: ($\sigma_{\rm WISE} = 0.686$ mag, $\sigma_{\rm
  K} = 0.681$ mag).
\end{enumerate}

\begin{figure}
\plotone{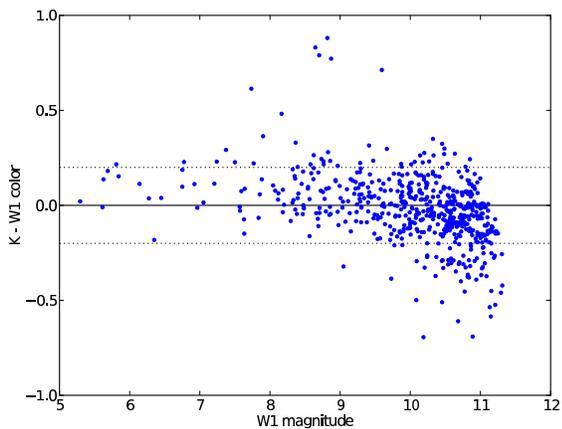}
\caption{$K - W1$ vs. $W1$ color-magnitude diagram of the {\it WISE}
  TF galaxy sample.  The ($K - W1$) = 0 color is represented by the
  solid line, while the dashed lines enclose ($K - W1$) = $\pm$ 0.2.
  Over 80\% of all {\it WISE} TF galaxies fall within this region,
  suggesting that light from the $K$-and $W1$-bands trace similar
  stellar populations (including older, more centrally-located stars.)
  It is likely that the similarity in color between the two bands
  helps to explain the excellent agreement between the {\it WISE} TF
  and 2MTF K-band relations. \label{fig:KW_colors}}
\end{figure}

This agreement can (at least partially) be explained by examining the
$(K - W1)$ colors of typical TF galaxies.  In Figure
\ref{fig:KW_colors} we show the color magnitude diagram for the
galaxies in the final {\it WISE} TF sample.  While there are a few
noticeable outliers, a large majority of objects fall within 0.5
magnitudes of $(K - W1) = 0$.  Of these, nearly all are also within
0.2 magnitudes, save for a slight blueward trend at $m_{W1} \geq
10.75$.  Given this near-neutral color, which suggests that the $K$-
and $W1$-bands trace similar stellar populations, (including the
older, centrally-located bulge stars near the cores of galaxies), it
is clear the features of TF relations constructed in these two bands
should very nearly mimic each other.

%--------------------------------------------------------------------
%--------------------------------------------------------------------
%--------------------------------------------------------------------

\begin{figure}
\plotone{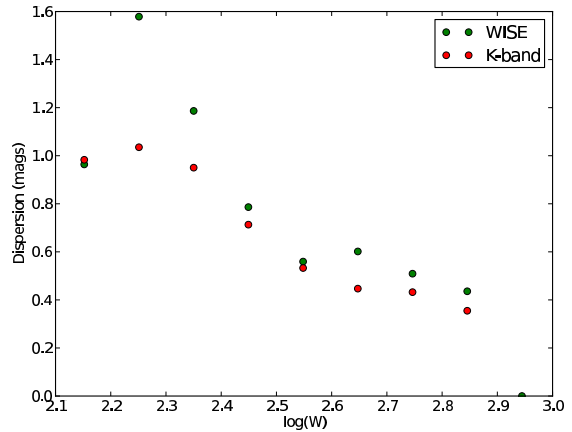}
\caption{Average scatter of the TF relation, binned as a function of
  line width.  We can see that the {\it WISE} TF relation (green
  points) systematically shows a larger scatter than the K-band TF
  relation (red points). However, both relations show an overall
  decrease in dispersion as line width
  decreases.\label{fig:dispersions}}
\end{figure}

Like the 2MTF relations, the total scatter of the {\it WISE} TF
relation appears to qualitatively decrease towards the bright end of
the sample.  Therefore, in addition to calculating a dispersion for
the full data set, we also measure dispersions for subsamples of the
data, binned by line width.  This result is shown in Figure
\ref{fig:dispersions}.  As expected, the {\it WISE} TF dispersion
(represented by green points) does indeed decrease as line width
increases, with narrow-end galaxies showing an average dispersion of
$\sim$ 1 magnitude, and wide-end galaxies showing an average
dispersion of $\sim$ 0.5 magnitudes.  Figure \ref{fig:dispersions}
also presents the binned K-band dispersion (red points).  Comparing
the two data sets, we see that the K-band relation is systematically
lower than the {\it WISE} TF, but by $\lesssim$ 0.1 magnitudes, which
is within the level of statistical uncertainty.

%--------------------------------------------------------------------
%--------------------------------------------------------------------

\subsection{Tully-Fisher and Bulge Light in the Mid-Infrared}
Large galaxies like M31, which can compose up to 5\% of a TF sample,
have a prominent bulge in the mid-IR, much more so than in the
optical.  Assuming Model A of \citet{wps03}, M31-like galaxies have a
bulge-to-disk luminosity ratio given by L$_R$(bulge)/L$_R$(disk) =
(2.5/2.7) $\times$ (4.4/7) = 0.58 in the $R$-band. If the bulge is a
magnitude redder than the disk in $(R - W1)$, the corresponding
bulge-to-disk ratio in the mid-IR will instead be
L$_{W1}$(bulge)/L$_{W1}$(disk) = 1.45.  This suggests that {\it WISE}
magnitudes of the largest galaxies tend to be bulge-dominated.

In addition to the luminosity, the presence of a bulge also affects
the dynamical parameter (i.e., line width) in the TF relation
(\citealt{ton11}). Switching a bulge on and off in a large galaxy,
then, could perturb such a galaxy's position in the TF relation by as
much as 0.1 dex. Bulge variance in massive galaxies therefore
contributes to the scatter in Tully-Fisher, and may be partially
responsible for the measured dispersion in the {\it WISE} TF relation.

%--------------------------------------------------------------------
\begin{figure*}
\begin{center}
\includegraphics[width=17.6cm]{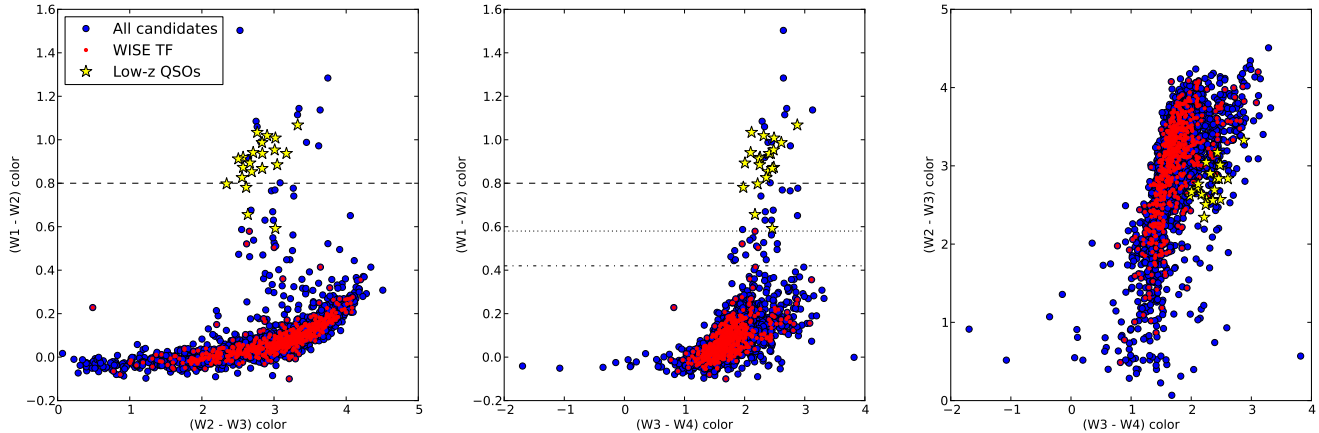}
\end{center}
\caption{Various {\it WISE} color-color diagrams used to identify and
  remove Active galaxies (AGN) from normal, star-forming TF galaxies.
  In each panel, we plot the final {\it WISE} TF galaxy sample (red
  points), the full 2MTF candidate catalog (blue circles) and a set of
  known $z < 0.1$ quasars (QSOs, yellow stars).  {\bf Left:} The
  ($W1-W2$)/($W2-W3$) color space used in \citet{yan13}.  We can see a
  clear division between the QSOs and TF galaxies along the ($W1-W2 $)
  axis, though there is a slight mixing between the QSOs and
  (rejected) 2MTF candidates.  The ($W1-W2 \geq 0.8$) color cut used
  by \citet{ste12} to select AGN is shown as a dashed line.  While
  this cut does cleanly remove many of the QSOs, some still survive,
  suggesting that this cut is too restrictive for our purposes.  {\bf
    Middle:} The ($W1-W2$)/($W3-W4$) color space.  Here again we see
  the division of QSOs and normal galaxies along the ($W1-W2$) axis,
  but the overall ``AGN locus'' is much more prominent, largely thanks
  to a greatly reduced spread along the ($W3-W4$) axis.  Along with
  the ($W1-W2 > 0.8$) line, we also plot the less restrictive ($W1-W2
  \geq 0.58$) color cut, which just separates the reddest {\it WISE}
  TF galaxy from the bluest QSO, as a dotted line.  An even bluer cut
  ($W1-W2 \geq 0.42$) that traces the very edge of the AGN locus is
  plotted as a dash-dotted line, but given that a few {\it WISE} TF
  galaxies are also eliminated with this cut, it may eliminate too
  much.  {\bf Right:} The ($W2-W3$)/($W3-W4$) color space.  We include
  this panel to highlight the importance of the ($W1-W2$) axis, as
  without this color, the QSOs and normal galaxies are not nearly as
  well separated, providing little discriminatory
  power.\label{fig:wise_colors}}
\end{figure*}
%--------------------------------------------------------------------

\subsection{Identifying AGN contamination with {\it WISE} colors}

Galaxies with an active nucleus (AGN) can also add to the scatter of
the TF relation, as the flux coming from the active core can often
dwarf the stellar emission, leading to a mismatch between a galaxy's
total magnitude and line width.  Since it is possible that up to 10\%
of all emission-line galaxies in the Local Universe host an AGN
\citep{bri04}, it is important to identify and eliminate these objects
from potential TF samples.  While there are a number of successful
techniques capable of identifying and segregating AGN from other
galaxies, one particularly apropos method utilizes a combination of
mid-IR colors \citep{lac04,ste05}.  Originally developed using the
\textit{Spitzer} IRAC wavebands, the concept has recently been
extended to include {\it WISE} photometry, where it has been shown
that a ($W1 - W2$) color selection is remarkably efficient at
segregating AGN from normal, star-forming galaxies -- at least for
objects with redshifts $z \leq 3$ \citep[see e.g.,
][]{ash09,aas10,eck10}.  \citet{ste12} and \citet{yan13} have further
included {\it WISE} $W3$- and $W4$-band information to the detection
efforts, allowing for an improved targeting of dust-obscured AGN as
well.

While all of these studies share the common goal of obtaining the
largest, cleanest sample of AGN, we seek to do the exact opposite --
namely, determine the {\it WISE} color parameter(s) that will most
efficiently \textit{eliminate} AGN while keeping as many normal, star
forming galaxies as possible.

In Figure \ref{fig:wise_colors}, we plot several color-color diagrams
of the {\it WISE} TF sample.  Along with the final {\it WISE} TF
galaxies (red points) we additionally plot the full 2MTF candidate
galaxy catalog (blue circles), in order to see if there are any
significant differences between ``good'' TF galaxies and those
rejected according to other physical parameters.  We also include a
set of known low-redshift ($z < 0.1$) AGN (yellow stars), taken from
the \citet{wu12} quasar (QSO) catalog, in order to better see the
differences between active and normal galaxies.

In ($W1-W2$)/($W2-W3$) color space (left panel), we see (like previous
studies) a clear separation between the QSO and TF galaxies, with the
QSOs having a significantly redder ($W1-W2$) color.  Intriguingly, the
QSO color space is also populated with several rejected TF galaxy
candidates, suggesting that our applied data cuts (Section \ref{data})
can at least partially eliminate some AGN contamination already.
While no single cut parameter eliminates all of the galaxies above the
QSO line, the line-width quality, inclination-angle, and morphology
cuts are responsible for removing $\sim$ 80\% of these objects.
Replacing the ($W2-W3$) color with ($W3-W4$) (middle panel), we still
see the prominent separation between the QSOs and TF galaxies, but the
overall AGN locus is much more pronounced, thanks to the reduced
spread of the ($W3-W4$) color.  In many ways the ``best'' dividing
line between QSOs and normal galaxies is much easier to identify in
this color space, and it may well be that, in the very local Universe,
the ($W1-W2$)/($W3-W4$) parameter is an even more efficient AGN/TF
discriminator.  For the 2MTF catalog, cutting galaxies with ($W1-W2$)
$> 0.58$ should cleanly remove bright AGN without significantly
impacting the galaxy sample.
Finally, to highlight the importance of the ($W1-W2$) color, we also
plot objects in ($W2-W3$)/($W3-W4$) color space (right panel).  Unlike
previous color spaces, this space does not significantly differentiate
between AGN and normal galaxies, severely limiting its usefulness as a
TF galaxy discriminator.

\section{Conclusions}

We have presented a mid-infrared extension of the Tully-Fisher
relation, using luminosities obtained from the {\it WISE} satellite.
The galaxies used in this work are taken from the all-sky,
near-infrared 2MTF galaxy catalog and cover distances out to redshift
$z \approx 0.045$.  Prior to constructing the full TF relation, we
apply a number of corrections to both the magnitudes and line widths
of these galaxies, in order to reduce the observed TF scatter and
better compare our results to previous work.  In particular we pay
special attention to correcting the effects of peculiar velocities,
using the model of \citet{erd06}, and we also apply a morphology
correction in order to normalize magnitudes to an Sc-type template.

After applying the corrections, we use a bivariate, least squares
fitting routine to generate the best-fit TF parameters, finding a
slope of $a_{\rm TF} = -10.05$ and an intercept of $b_{\rm TF} =
2.89$.  Transforming the line widths into a magnitude measurement
leads to the final {\it WISE} TF relation, given by:
\begin{equation}
M_{\rm corr} = -22.24 - 10.05 [\log(W_{\rm corr}) - 2.5]
\end{equation}
We calculate the dispersion around the best fit model, measuring an
average {\it WISE} TF scatter of $\sigma_{\rm WISE} = 0.686$.
However, we find that the overall scatter is strongly dependent on
line width, with galaxies near $\log(W_{\rm corr}) = 2.3$ having a
dispersion approximately 0.5 mags greater than galaxies near
$\log(W_{\rm corr}) = 2.9$.

We find that the {\it WISE} TF relation is significantly different
from the preliminary 3.6 $\mu$m \textit{Spitzer} TF relation presented
in \citet{fre10}, with {\it WISE} TF having both a steeper slope and
larger scatter.  This is not surprising, however, given the small
sample size and unusually bright galaxies chosen in that work.
Conversely, we find good agreement between {\it WISE} TF and the
cluster-based mid-IR TF relation presented in \citet{sor13}.  Given
these similarities, in spite of somewhat different analytical
techniques, we conclude that a steeper, more scattered TF relation
better describes the properties of galaxies observed in the mid-IR
than the optimistic results of \citet{fre10}.

Finally, we also compare {\it WISE} TF to the near-IR ($K$-band) 2MTF
template relation presented in \citet{mas08}, and find that, both in
terms of TF parameters and in total dispersion, the two relations
agree extremely well, with both showing a TF slope near the
theoretical limit of $L \propto v_{\rm rot}^4$ predicted from physical
arguments.  This result may not be entirely surprising, given that the
typical $(K - W1)$ color of {\it WISE} TF galaxies is $\sim$ 0, and it
suggests that the $K$-band and $W1$-bands are tracing similar stellar
populations (including the older, centrally-located stars in the
bulge).  Whether this agreement holds for TF relations constructed at
longer wavelengths -- where the observed light from thermally-heated
dust will ostensibly begin to trace the younger, disk-dominated stars
again -- is unknown, but would be an interesting question to examine
in future work.

\begin{acknowledgments}

The authors wish to thank the anonymous referee for thoroughly reading
this manuscript and for providing several helpful comments and
suggestions that greatly improved the work.  We also thank Pirin
Erdo{\u g}du for providing the initial peculiar velocity model.  This
research was conducted by the Australian Research Council Centre of
Excellence for All-sky Astrophysics (CAASTRO), through project number
CE110001020.  This publication makes use of data products from the
Wide-field Infrared Survey Explorer, which is a joint project of the
University of California, Los Angeles, and the Jet Propulsion
Laboratory/California Institute of Technology, funded by the National
Aeronautics and Space Administration.  This research has made use of
the NASA/IPAC Extragalactic Database (NED) which is operated by the
Jet Propulsion Laboratory, California Institute of Technology, under
contract with the National Aeronautics and Space Administration.

\end{acknowledgments}

\begin{appendix}
\section{Testing the accuracy of the peculiar velocity interpolation}
While the interpolation procedure we use to smooth the \citet{erd06}
peculiar velocity model allows us to select a more diverse range of
peculiar velocities, these values may not always be correct,
especially if the real velocity field varies in an unexpected way
between data points.  Inaccurate velocity corrections can cause
problems in the analysis, fundamentally change the fitted TF
parameters, or at the very least increase the overall TF dispersion,
leading to a pessimistic assessment of {\it WISE} magnitudes for TF
purposes.  Therefore, to check the accuracy of our interpolation
scheme we compare our model peculiar velocities to real data, using
galaxies with known redshifts and (redshift-independent) distances.

To collect a sample of galaxies with known distances, we turn to the
NED-D
database\footnote{http://ned.ipac.caltech.edu/Library/Distances/}, an
online compilation of redshift-independent distance measurements taken
from the literature.  While the full NED-D catalog contains over 60000
measurements, we limit our selection to distances calculated from
Type-Ia supernova data, and further limit this subsample to galaxies
located less than 247.5 Mpc away from the Earth (the limit of the
\citealt{erd06} model).  This leaves us with a total of 2189
measurements for 373 unique galaxies.

We calculate peculiar velocities from these data using the standard
technique of combining the distance measurement ($D_{\rm obs}$) with its
assumed Hubble constant ($H_0$, also provided from NED-D) and redshift
($z_{\rm obs}$), according to:
\begin{equation}
v_{\rm pec, ~obs} = c \cdot z_{\rm obs} - H_0 \cdot D_{\rm obs}
 ~~\equiv~~ v_{\rm T, obs} - v_{\rm H, obs}
\end{equation}
where $v_{\rm T}$ and $v_{\rm H}$ are, respectively, the total and
Hubble flow recessional velocities mentioned in Section \ref{vpec}.

For each observed peculiar velocity, we also generate a corresponding
model value using our interpolation method.  Since we already know the
distances to these objects (and hence, their true position in model
space), we do not need to iterate over the total velocity field like
we did for the galaxies in the main TF sample (Section \ref{vpec}) to
measure a model peculiar velocity.  Instead, the peculiar velocity
can be measured directly, using an object's supergalactic coordinates
and observed distance:
\begin{equation}
v_{\rm pec, ~mod} = v_{\rm P}(sgl,sgb,D_{\rm obs})
\end{equation}
where $v_{\rm P}$ is simply the \citet{erd06} model without a Hubble
flow component.

%----------------------------------------------------------
\begin{figure}
\plottwo{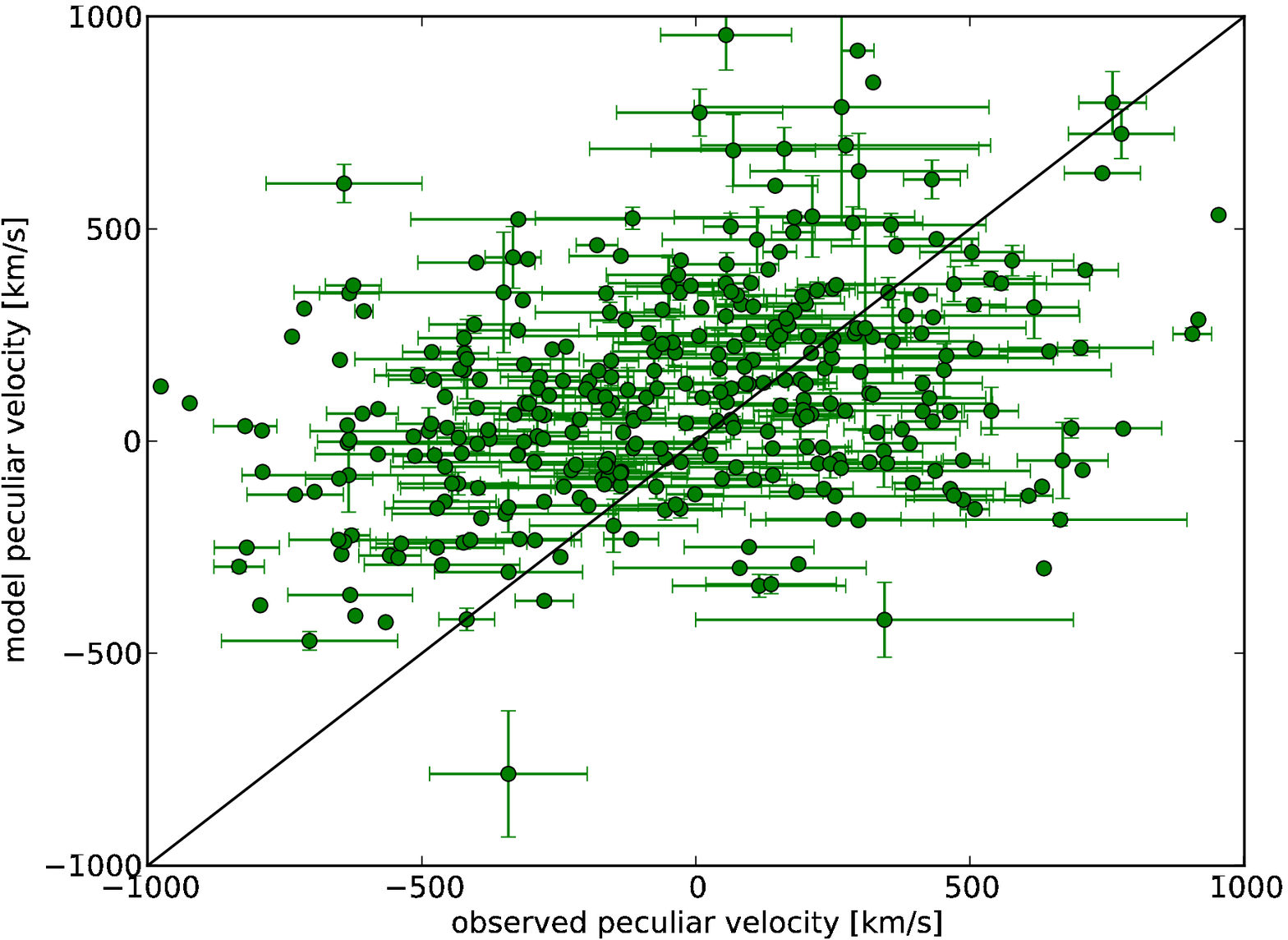}{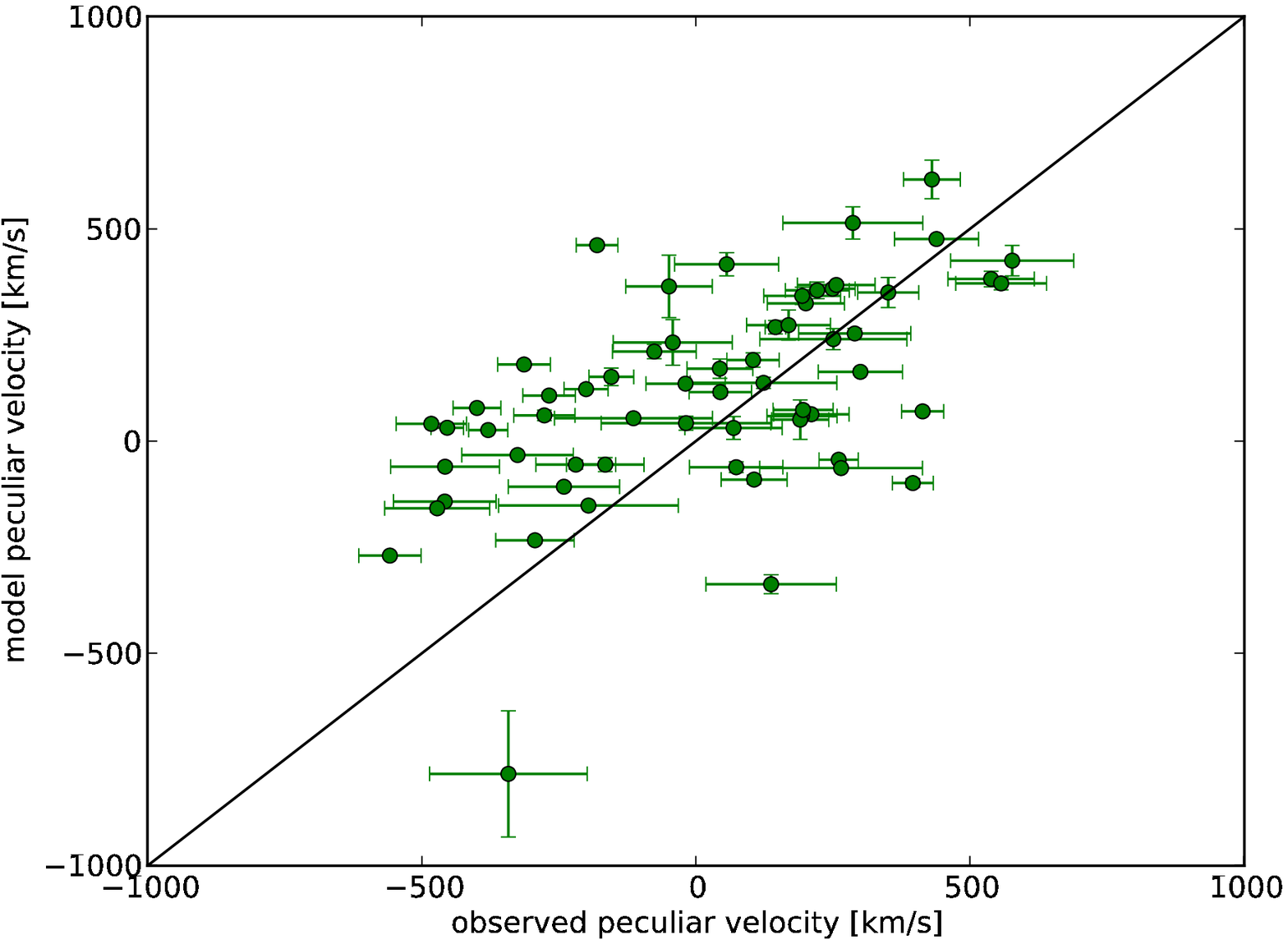}
\caption{Comparisons between observed and model peculiar velocities,
  using galaxies with known redshift-independent distance
  measurements.  {\bf Left:} All galaxies with Type-Ia supernova
  distance measurements from the on-line NED-D database, within a
  247.5 Mpc radius of the Earth.  {\bf Right:} Only those galaxies in
  the database with at least 10 independent distance measurements.  In
  both panels, any galaxy with multiple distance measurements is
  represented by a single (averaged) data point, with error bars
  representing the dispersions on the
  measurements. \label{fig:vpec_comp}}
\end{figure}
%----------------------------------------------------------

The comparison between observed and model peculiar velocity values is
presented in Figure \ref{fig:vpec_comp}.  The left panel of the figure
shows all galaxies taken from the NED-D catalog, while the right panel
shows only those galaxies which have at least 10 independent distance
measurements.  In both panels (for the sake of clarity) we average all
measurements of a given galaxy into a single data point, using error
bars to represent the dispersion of both quantities.  

Overall, a meaningful description of the differences between the
observed and model peculiar velocities is difficult, especially given
that the error bars on the observed velocities are large.  While the
full galaxy sample shows a mild positive correlation between the two
quantities ($\rho_{\rm (full)} = 0.30$), there is still a significant
dispersion around the 1:1 correspondence line ($\sigma_{\rm (full)} =
350$ km s$^{-1}$).  However, the correlation increases ($\rho_{\rm
  (m>10)} = 0.57$) and the dispersion decreases ($\sigma_{\rm (m>10)}
= 300$ km s$^{-1}$) when focusing only on the galaxies with 10 or more
distance measurements, suggesting that, at least in cases where the
distance to a galaxy is well constrained, there is not as significant
a disagreement between observation and our model.

Given the moderate success in recreating observed peculiar velocities,
then, we believe that our interpolated peculiar velocity model will
not statistically bias the magnitudes of our galaxy sample, and can
thus be safely used to construct the {\it WISE} TF relation.

\section{Measuring the effects of the peculiar velocity model}
\begin{figure}
\plottwo{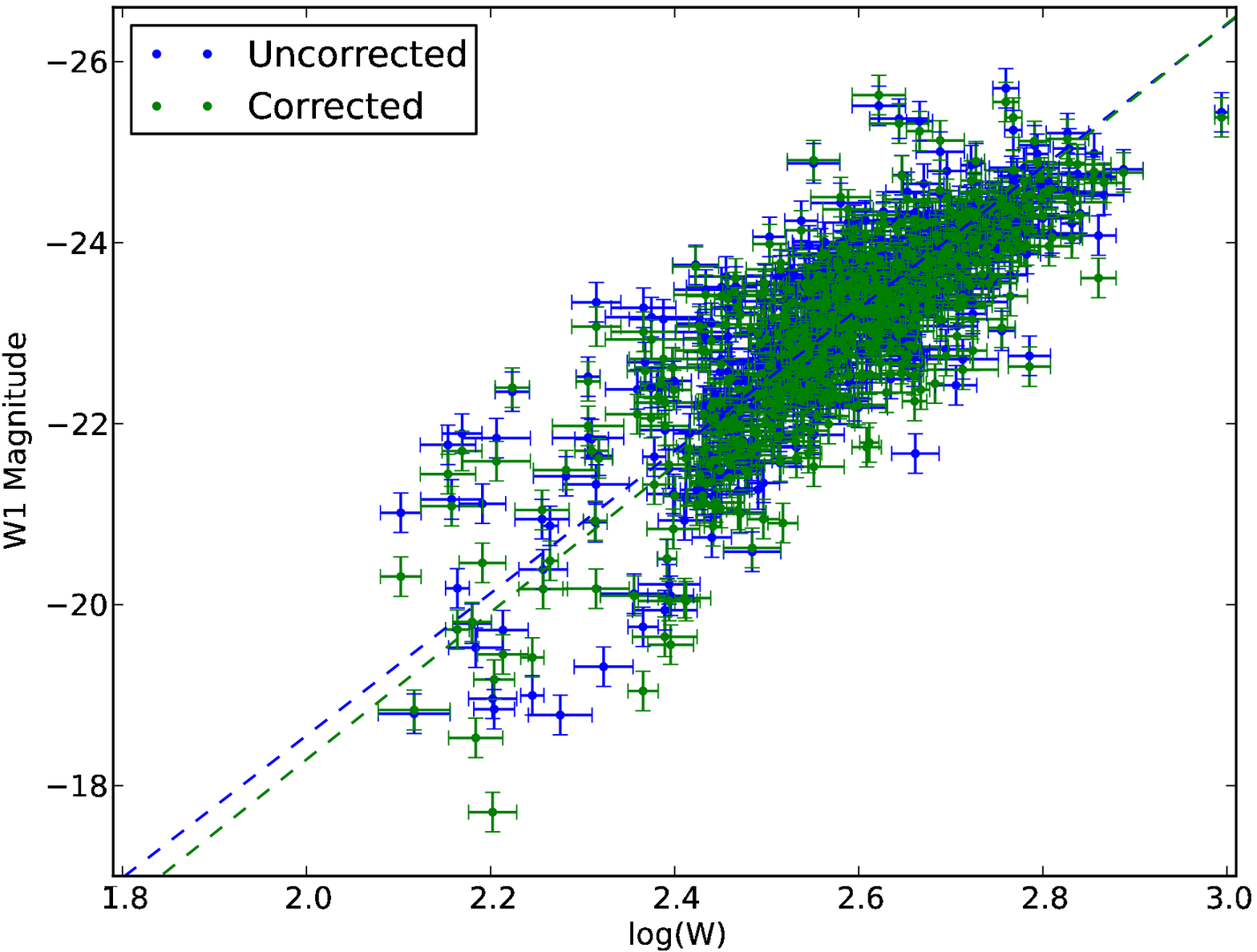}{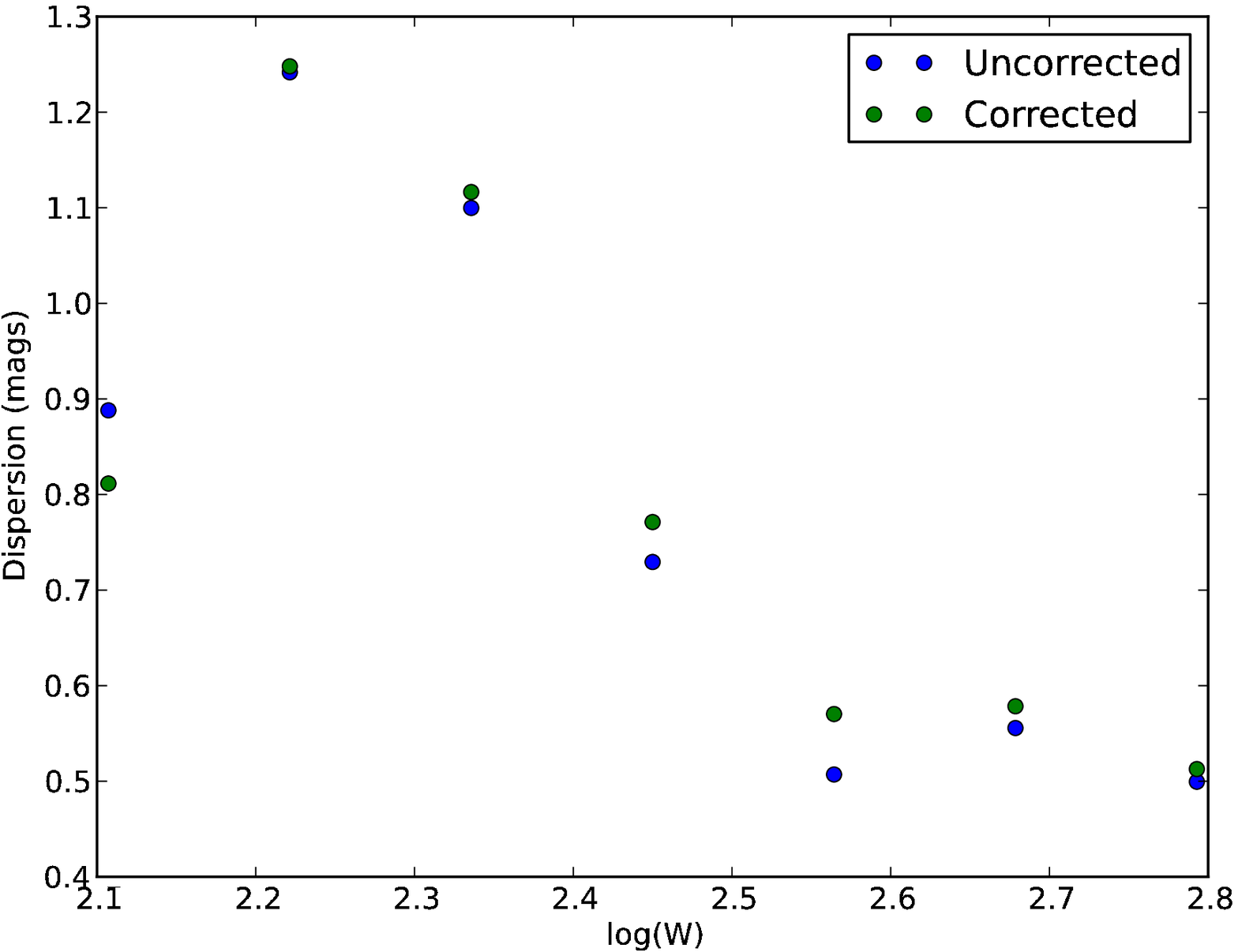}
\caption{Effects of the peculiar velocity correction on the {\it WISE}
  TF relation.  {\bf Left:} The full galaxy sample and best-fit TF
  lines.  {\bf Right:} Measured dispersion around the best-fit TF
  parameters.  In both the overall appearance and total dispersion the
  peculiar-velocity corrected TF relation (green) does not
  significantly deviate from the uncorrected (blue)
  relation.\label{fig:vpec_results}}
\end{figure}

To observe the effects of peculiar velocity on the TF relation, we
rerun the analysis without applying the interpolated model, then
compare the parameters of this new TF model to the original.  This is
done after applying the initial magnitude and line width corrections
(Section \ref{initial_TF}), but \textit{before} applying the
morphology correction (Section \ref{morphology}).  Figure
\ref{fig:vpec_results} shows the results of this comparison, both in
terms of the fit parameters (left panel) and total dispersion (right
panel).  Overall, we find that the peculiar velocity correction does
not significantly alter the final results.  While the slope of the
uncorrected relation is nominally shallower ($a_{\rm uncorr} =
-7.86$), the intercept is brighter ($b_{\rm uncorr} = -2.83$) and the
dispersion measurements differ (the corrected relation shows slightly
lower dispersions at the narrow line width end and slightly higher
dispersions at the wide end), all of these changes are well within the
current limits of statistical uncertainty.  This is not entirely
unexpected, given that on average a peculiar velocity correction
alters galaxy magnitudes by less than 0.2 magnitudes, while the
measured dispersion of the overall {\it WISE} TF relation ranges
between 0.5 and 1.5 magnitudes (Figure \ref{fig:dispersions}).

This suggests that peculiar velocities do not contribute significantly
to the dispersion budget of the TF relation (at least when dealing
with mid-IR magnitudes), and that the scatter is instead driven by
some other, unidentified parameter.  However, even though the total
effect is small, we still choose to include the peculiar velocity
correction in our analysis, for the sake of completeness.

%----------------------------------------------------------
%----------------------------------------------------------

\end{appendix}

%\newpage

\begin{turnpage}
\begin{deluxetable}{lrrrcrrrrrrrrrrrrr}
\tablecaption{Tully-Fisher Data. \label{tbl:TF_data}}
\tablehead{
\colhead{Galaxy Name} & \colhead{RA} & \colhead{Dec} & \colhead{$\theta_{i}$} & \colhead{T-Type} & \colhead{$v_{\rm obs}$} & \colhead{$v_{\rm H}$} & \colhead{$m_{\rm obs}$} & \colhead{$\mu_{\rm H}$} & \colhead{$m_{\rm Int}$} & \colhead{$A_{3.6}$} & \colhead{$\Delta M$} & \colhead{$M_{\rm corr}$} & \colhead{$M_{\rm err}$} & \colhead{$W_{\rm IC}$} & \colhead{$W_{\rm corr}$} & \colhead{$W_{\rm err}$} & \colhead{clip}\\
\colhead{(1)} & \colhead{(2)} & \colhead{(3)} & \colhead{(4)} & \colhead{(5)} & \colhead{(6)} & \colhead{(7)} & \colhead{(8)} & \colhead{(9)} & \colhead{(10)} & \colhead{(11)} & \colhead{(12)} & \colhead{(13)} & \colhead{(14)} & \colhead{(15)} & \colhead{(16)} & \colhead{(17)} & \colhead{(18)}
}
\startdata
UGC 00005  &   0.773583  &  -1.913798  &60.9  &4  &7288  &7384  &10.167  &-34.341  & 0.008  & 0.051  & 0.000  &-24.233  & 0.218  &461  &520  &23  &N  \\
NGC 7814  &   0.812280  &  16.145229  &68.0  &2  &1051  &1063  & 7.053  &-30.132  & 0.009  & 0.064  & 0.239  &-23.391  & 0.217  &460  &489  &9  &N  \\
UGC 00014  &   0.895900  &  23.200779  &46.8  &4  &7254  &7206  &10.767  &-34.289  & 0.019  & 0.027  & 0.000  &-23.567  & 0.219  &332  &446  &34  &N  \\
UGC 00039  &   1.392929  &  53.631325  &60.9  &5  &9523  &9513  &10.606  &-34.892  & 0.082  & 0.051  & 0.000  &-24.418  & 0.218  &496  &560  &23  &N  \\
UGC 00040  &   1.451719  &  27.449430  &51.9  &3  &7526  &7602  &10.960  &-34.405  & 0.014  & 0.033  & 0.140  &-23.632  & 0.218  &362  &452  &26  &N  \\
UGC 00051  &   1.668191  &   5.113459  &48.7  &4  &5371  &5078  &11.306  &-33.529  & 0.006  & 0.030  & 0.000  &-22.258  & 0.219  &280  &364  &24  &N  \\
NGC 0002  &   1.821281  &  27.678375  &54.7  &2  &7547  &7637  &11.070  &-34.415  & 0.013  & 0.037  & 0.032  &-23.426  & 0.218  &344  &413  &19  &N  \\
NGC 0010  &   2.143900  & -33.858337  &60.9  &6  &6804  &6515  & 9.236  &-34.070  & 0.002  & 0.051  & 0.000  &-24.887  & 0.217  &550  &622  &17  &N  \\
NGC 0023  &   2.472548  &  25.923786  &53.3  &1  &4565  &4164  & 8.689  &-33.098  & 0.007  & 0.035  & 0.337  &-24.788  & 0.218  &382  &468  &24  &N  \\
NGC 0019  &   2.670312  &  32.983078  &60.9  &4  &4788  &4316  &10.512  &-33.175  & 0.008  & 0.051  & 0.000  &-22.722  & 0.218  &302  &338  &10  &N  \\

\tablenotetext{}{Identifying features and physical parameters for each
  galaxy in the master {\it WISE} TF relation.  In column 1, we
  present the galaxy's preferred common name, as displayed in NED.
  Columns 2 and 3 show galaxy coordinates in decimal degrees.  The
  inclination angle (also in degrees) is shown in column 4.  Galaxy
  morphology is presented in column 5, in the form of the de
  Vaucouleurs T-type classification scheme.  The observed and
  Hubble-flow recessional velocities (in km s$^{-1}$), are displayed
  in columns 6 and 7, respectively.  The observed, apparent {\it WISE}
  W1 magnitude is presented in column 8, and the distance modulus is
  presented in column 9.  Corrections to the observed galaxy magnitude
  are displayed in columns 10 through 12, with column 10 showing the
  correction for internal extinction, column 11 showing the correction
  for Galactic extinction, and column 12 showing any morphology-based
  corrections, normalizing all galaxies to type Sc (T $= 5$).  The
  final, fully corrected absolute magnitude is presented in column 13,
  and its error is shown in column 14.  The observed
  (instrument-corrected) line width (in km s$^{-1}$) is shown in
  column 15, while the fully-corrected line width is presented in
  column 16, along with its uncertainty in column 17.  Finally column
  18 shows the sigma-clipping flag for each galaxy, with 'N' meaning
  the galaxy is included in the final TF fit, and 'Y' meaning the
  galaxy is removed prior to the final fit.}

\tablenotetext{}{~}

\tablenotetext{}{Table \ref{tbl:TF_data} is published in its entirety
  in the electronic edition of {\it The Astrophysical Journal}. A
  portion is shown here for guidance regarding its form and content.}

\end{deluxetable}
\end{turnpage}


\begin{thebibliography}{}
\bibitem[Aaronson et al.(1979)]{aar79} Aaronson, M., Huchra, J., \&
  Mould, J.\ 1979, \apj, 229, 1
\bibitem[Aaronson et al.(1980)]{aar80} Aaronson, M., Mould, 
J., \& Huchra, J.\ 1980, \apj, 237, 655 
\bibitem[Ashby et al.(2009)]{ash09} Ashby, M.~L.~N., Stern, D.,
  Brodwin, M., et al.\ 2009, \apj, 701, 428
\bibitem[Assef et al.(2010)]{aas10} Assef, R.~J., Kochanek, C.~S.,
  Brodwin, M., et al.\ 2010, \apj, 713, 970
\bibitem[Bernstein et al.(1994)]{ber94} Bernstein, G.~M.,
  Guhathakurta, P., Raychaudhury, S., et al.\ 1994, \aj, 107, 1962
\bibitem[Bothun \& Mould(1987)]{bot87} Bothun, G.~D., \& Mould,
  J.~R.\ 1987, \apj, 313, 629
\bibitem[Brinchmann et al.(2004)]{bri04} Brinchmann, J., Charlot, S.,
  White, S.~D.~M., et al.\ 2004, \mnras, 351, 1151
\bibitem[Buta et al.(2010)]{but10} Buta, R.~J., Sheth, K., Regan, M.,
  et al.\ 2010, \apjs, 190, 147
\bibitem[Dale et al.(2007)]{dal07} Dale, D.~A., Gil de Paz, A.,
  Gordon, K.~D., et al.\ 2007, \apj, 655, 863
\bibitem[Davis \& Peebles(1983)]{dav83} Davis, M., \& Peebles,
  P.~J.~E.\ 1983, \araa, 21, 109
\bibitem[Djorgovski \& Davis(1987)]{djo87} Djorgovski, S., \& Davis,
  M.\ 1987, \apj, 313, 59
\bibitem[Dressler(1987)]{dressler87} Dressler, A.\ 1987, \apj, 317, 1
\bibitem[Dressler et al.(1987)]{dre87} Dressler, A., Lynden-Bell, D.,
  Burstein, D., et al.\ 1987, \apj, 313, 42
\bibitem[Eckart et al.(2010)]{eck10} Eckart, M.~E., McGreer, I.~D.,
  Stern, D., Harrison, F.~A., \& Helfand, D.~J.\ 2010, \apj, 708, 584
\bibitem[Erdo{\u g}du et al.(2006)]{erd06} Erdo{\u g}du, P., Lahav,
  O., Huchra, J.~P., et al.\ 2006, \mnras, 373, 45
\bibitem[Faber \& Jackson(1976)]{fj76} Faber, S.~M., \& Jackson,
  R.~E.\ 1976, \apj, 204, 668
\bibitem[Ferrarese \& Merritt(2000)]{fer00} Ferrarese, L., \& Merritt,
  D.\ 2000, \apjl, 539, L9
\bibitem[Freedman \& Madore(2010)]{fre10} Freedman, W.~L., \& Madore,
  B.~F.\ 2010, \araa, 48, 673
\bibitem[Giovanelli et al.(1994)]{gio94} Giovanelli, R., Haynes,
  M.~P., Salzer, J.~J., et al.\ 1994, \aj, 107, 2036
\bibitem[Giovanelli et al.(1997)]{gio97} Giovanelli, R., Haynes,
  M.~P., Herter, T., et al.\ 1997, \aj, 113, 53
\bibitem[Hong et al.(2013)]{hon13} Hong, T., Staveley-Smith, L.,
  Masters, K.~L., et al.\ 2013, arXiv:1304.0882
\bibitem[Huchra et al.(2012)]{huc12} Huchra, J.~P., Macri, L.~M.,
  Masters, K.~L., et al.\ 2012, \apjs, 199, 26
\bibitem[Jarrett et al.(2012)]{jar12} Jarrett, T.~H., Masci, F., Tsai,
  C.~W., et al.\ 2012, \aj, 144, 68
\bibitem[Jarrett et al.(2013)]{jar13} Jarrett, T.~H., Masci, F., Tsai,
  C.~W., et al.\ 2013, \aj, 145, 6
\bibitem[Lacy et al.(2004)]{lac04} Lacy, M., Storrie-Lombardi, L.~J.,
  Sajina, A., et al.\ 2004, \apjs, 154, 166
\bibitem[Macri(2001)]{mac01} Macri, L.~M.\ 2001, Ph.D.~Thesis,
\bibitem[Masters et al.(2003)]{mas03} Masters, K.~L., Giovanelli, R.,
  \& Haynes, M.~P.\ 2003, \aj, 126, 158
\bibitem[Masters et al.(2006)]{mas06} Masters, K.~L., Springob, C.~M.,
  Haynes, M.~P., \& Giovanelli, R.\ 2006, \apj, 653, 861
\bibitem[Masters et al.(2008)]{mas08} Masters, K.~L., Springob, C.~M.,
  \& Huchra, J.~P.\ 2008, \aj, 135, 1738
\bibitem[Marinoni et al.(1998)]{mar98} Marinoni, C., Monaco, P.,
  Giuricin, G., \& Costantini, B.\ 1998, \apj, 505, 484
\bibitem[Mocz et al.(2012)]{moc12} Mocz, P., Green, A., Malacari, M.,
  \& Glazebrook, K.\ 2012, \mnras, 425, 296
\bibitem[Paturel et al.(2003)]{pat03} Paturel, G., Theureau, G.,
  Bottinelli, L., et al.\ 2003, \aap, 412, 57
\bibitem[Peebles(1976)]{pee76} Peebles, P.~J.~E.\ 1976, \apss, 45, 3
\bibitem[P{\'e}rez et al.(2013)]{per13} P{\'e}rez, E., Cid Fernandes,
  R., Gonz{\'a}lez Delgado, R.~M., et al.\ 2013, \apjl, 764, L1
\bibitem[Rothberg et al.(2000)]{rot00} Rothberg, B., Saunders, W.,
  Tully, R.~B., \& Witchalls, P.~L.\ 2000, \apj, 533, 781
\bibitem[Sakai et al.(2000)]{sak00} Sakai, S., Mould, J.~R., Hughes,
  S.~M.~G., et al.\ 2000, \apj, 529, 698
\bibitem[Schlegel et al.(1998)]{sch98} Schlegel, D.~J., Finkbeiner,
  D.~P., \& Davis, M.\ 1998, \apj, 500, 525
\bibitem[Skrutskie et al.(2006)]{skr06} Skrutskie, M.~F., Cutri,
  R.~M., Stiening, R., et al.\ 2006, \aj, 131, 1163
\bibitem[Sorce et al.(2012a)]{sor12a} Sorce, J.~G., Courtois, H.~M., \&
  Tully, R.~B.\ 2012, \aj, 144, 133
\bibitem[Sorce et al.(2012b)]{sor12b} Sorce, J.~G., Tully, R.~B., \&
  Courtois, H.~M.\ 2012, \apjl, 758, L12
%\bibitem[Sorce et al.(2013)]{sor13} Sorce, J.~G., Courtois, H.~M.,
%  Tully, R.~B., et al.\ 2013, arXiv:1301.4833
\bibitem[Sorce et al.(2013)]{sor13} Sorce, J.~G., Courtois, H.~M.,
  Tully, R.~B., et al.\ 2013, \apj, 765, 94
\bibitem[Springob et al.(2005)]{spr05} Springob, C.~M., Haynes, M.~P.,
  Giovanelli, R., \& Kent, B.~R.\ 2005, \apjs, 160, 149
\bibitem[Stern et al.(2005)]{ste05} Stern, D., Eisenhardt, P.,
  Gorjian, V., et al.\ 2005, \apj, 631, 163
\bibitem[Stern et al.(2012)]{ste12} Stern, D., Assef, R.~J., Benford,
  D.~J., et al.\ 2012, \apj, 753, 30
\bibitem[Theureau et al.(1998)]{the98} Theureau, G., Bottinelli, L.,
  Coudreau-Durand, N., et al.\ 1998, \aaps, 130, 333
\bibitem[Theureau et al.(2005)]{the05} Theureau, G., Coudreau, N.,
  Hallet, N., et al.\ 2005, \aap, 430, 373
\bibitem[Theureau et al.(2007)]{the07} Theureau, G., Hanski, M.~O.,
  Coudreau, N., Hallet, N., \& Martin, J.-M.\ 2007, \aap, 465, 71
\bibitem[Tonini et al.(2011)]{ton11} Tonini, C., Maraston, C.,
  Ziegler, B., et al.\ 2011, \mnras, 415, 811
\bibitem[Tully \& Fisher(1977)]{tf77} Tully, R.~B., \& Fisher,
  J.~R.\ 1977, \aap, 54, 661
\bibitem[Tully \& Pierce(2000)]{tul00} Tully, R.~B., \& Pierce,
  M.~J.\ 2000, \apj, 533, 744
\bibitem[Tully et al.(1998)]{tul98} Tully, R.~B., Pierce, M.~J.,
  Huang, J.-S., et al.\ 1998, \aj, 115, 2264
\bibitem[Verheijen(2001)]{ver01} Verheijen, M.~A.~W.\ 2001, \apj, 563,
  694
\bibitem[Widrow et al.(2003)]{wps03} Widrow, L.~M., Perrett, K.~M., \&
  Suyu, S.~H.\ 2003, \apj, 588, 311
\bibitem[Wright et al.(2010)]{wri10} Wright, E.~L., Eisenhardt,
  P.~R.~M., Mainzer, A.~K., et al.\ 2010, \aj, 140, 1868
\bibitem[Wu et al.(2012)]{wu12} Wu, X.-B., Hao, G., Jia, Z., 
Zhang, Y., \& Peng, N.\ 2012, \aj, 144, 49 
\bibitem[Yan et al.(2013)]{yan13} Yan, L., Donoso, E., Tsai, C.-W., et
  al.\ 2013, \aj, 145, 55 
\bibitem[Zheng et al.(2007)]{zhe07} Zheng, X.~Z., Dole, H., Bell,
  E.~F., et al.\ 2007, \apj, 670, 301

\end{thebibliography}
\end{document}